\documentclass[twocolumn,aps,prl,longbibliography]{revtex4-1}
\usepackage{graphicx}
\usepackage{amsfonts}
\usepackage{amsmath}
\usepackage{amssymb}
\usepackage{bm}
\usepackage{hyperref}
\usepackage{slashed}
\begin{document}
\title{A renormalization group approach to non-Hermitian topological quantum criticality}
\author{Boran Zhou}
\affiliation{Department of Physics, Nanjing University, Nanjing 210093, China}
\author{Rui Wang}
\email{rwang89@nju.edu.cn}
\affiliation{Department of Physics, Nanjing University, Nanjing 210093, China}
\affiliation{National Laboratory of Solid State Microstructures and Collaborative Innovation Center of Advanced Microstructures, Nanjing University, Nanjing 210093, China}
\author{Baigeng Wang}
\email{bgwang@nju.edu.cn}
\affiliation{Department of Physics, Nanjing University, Nanjing 210093, China}
\affiliation{National Laboratory of Solid State Microstructures and Collaborative Innovation Center of Advanced Microstructures, Nanjing University, Nanjing 210093, China}

\begin{abstract}
Critical transition points between symmetry-broken phases are characterized as fixed points in the  renormalization group (RG) theory. We show that, following the standard Wilsonian procedure that traces out the large momentum modes, this well known fact can break down in non-Hermitian systems. Based on non-Hermitian Su-Schrieffer-Hegger (SSH)-type models, we propose a real-space decimation scheme to study the criticality between the topological and trivial phase. We provide concrete examples and an analytic proof to show that the real-space scheme perfectly overcomes the insufficiency of the standard method, especially in the sense that it always preserves the system at criticality as fixed points under RG. The proposed method can also greatly simplify the search of critical points for complicated non-Hermitian models by ruling out the irrelevant operators. These results pave the way towards more advanced RG-based techniques for the interacting non-Hermitian quantum systems.

\end{abstract}

%\pacs{}
\maketitle
\emph{Introduction.}--The theory of RG achieved enormous success on the critical phenomena and  phase transitions accompanied by symmetry breaking. A key observation that evokes fundamental  developments of RG is that, the correlation length $\xi$ emergent from Gaussian approximation is divergent and responsible for the singularities in various of thermal dynamical functions at the criticality \cite{book1}. Thus, the critical behavior is dominated by fluctuations that are statistically self-similar within the length scale $\xi$, ensuring that the critical point must be a fixed point under the dilation of short-distance fluctuations.  This invariance of critical point is a fundamental requirement, which is also obeyed by transitions between phases characterized by distinct topological numbers \cite{article2,article3}.
%The critical point  is usually gapless and enjoys divergent correlation length $\xi$ \cite{book1}.
Moreover, since the topological numbers, being integers robust against the continuous perturbation, cannot flow under dilation, RG trajectories cannot cross the critical point from one phase to another.
\begin{figure}[h]
\includegraphics[width=\linewidth]{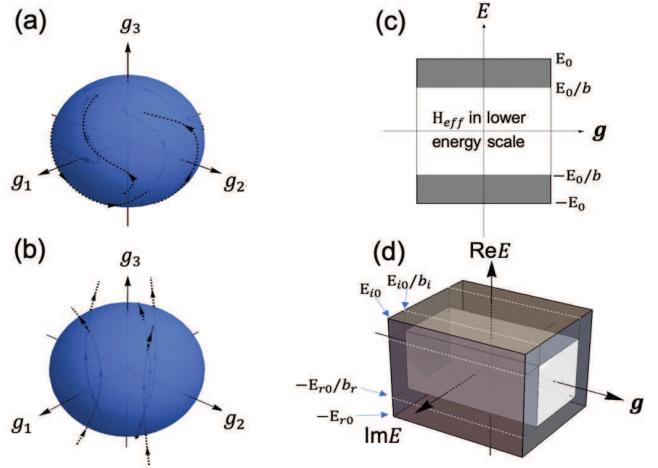}\label{fig1}
\caption{(color online).  Schematic plot of RG flow and the RG procedure for Hermitian and non-Hermitian systems. (a) and (b) indicate the CHS ($S^2$) and the RG flows in the Hermitian and non-Hermitian model, respectively. The flow generated by standard coarse-graining in momentum space remains within the CHS but flows out of the CHS in the former and latter case respectively. (c) The standard RG that integrates out the higher energy window, leading to an effective Hamiltonian in lower energy scale. (d) Both the real and imaginary window need to be decimated in non-Hermitian system, where the ratio of decimation is important to preserve the criticality.}
\end{figure}

Recent years also witnessed developments on non-Hermitian topological systems. The non-Hermicity, which can originate either from the loss and gain \cite{article4,article5,article6,article7} or the decaying quasi-particles in open systems \cite{article8,article9,article10,article11,article12,article13}, bring to forth a number of remarkable phenomena when it meets the nontrivial topology, including the breakdown of bulk-boundary correspondence \cite{article14,article15,article16,article17,article18,article19,article20,article21,article22,article23,article24,article25,article27,article28,article66},  the emergence of new topological invariants \cite{article16,article33,article17,article19,article34,article35,article36,article37,article38,article39,article40,article41}, and the non-Hermitian skin effect \cite{article42,article48,article49,article50,article51,article52,article53}. These results further lead to the born of the non-Bloch topological band theory, the enriched topological classifications according to symmetries \cite{article29,article30,article31}, new universal behaviors in conformal field theory \cite{article32}, and more \cite{article54,article57,article59,article60,article62,article65,article67,article68,article69,article70,article75,article77,article85,article87,article88,article89,article90}. On the other hand, the quantum phase transitions and the critical phenomena are much less studied in non-Hermitian topological systems. The RG approach that properly incorporates the non-Hermicity is yet to be investigated carefully.

In this work, we propose a RG method applicable to non-Hermitian topological systems.
%Our first  observation is that the standard Wilsonian RG scheme, where the fast mode in momentum space is decimated \cite{article55}, violates the requirement that the critical point must be a fixed point.
For purpose of clarity, we firstly introduce the general idea and results as follows. Assuming that a general model has N tuning parameters forming a vector, $\mathbf{g}=\{g_1,g_2,,...,g_N\}$, then,  in the phase diagram, the boundary separating different phases can in general be a manifold embedded in  $\mathbb{R}^N$, which we term as the critical hyper surface (CHS) for brevity, as schematically exemplified  by the $S^2$ embedeed in $\mathbb{R}^3$ in Fig.1(a),(b). The correlation function, as discussed before, is divergent in the manifold, therefore the RG transformation must keep the CHS invariant, namely, any RG flows with initial points located in the CHS should always remain within it. To obtain the flows, the coarse-graining is a standard way that traces out the high-energy degrees of freedom. In contrary to the Hermitain case where the modes with $|E|\in[E_0/b, E_0]$ ($b>1$) are eliminated (Fig.1(c)), there occurs two scaling parameters in non-Hermitian systems  $b_i$ and $b_r$, corresponding to the decimation of the real and imaginary window respectively, as shown by Fig.1(d). Intuitively, one expects that, unlike the Hermitian model where the flows always remain in CHS as indicated by Fig.1(a),  the ratio $b_i/b_r$ cannot be chosen arbitrarily in the non-Hermitian case, otherwise the CHS may not remain invariant as shown by Fig.1(b). This poses a fundamental question, i.e., the standard coarse-graining does not ensure the critical points to be fixed points with non-Hermicity. To overcome this problem, we propose a block-decimation RG scheme.  Using the non-Hermitian Su-Schrieffer-Heeger (SSH)-type models as examples, we show that, the proposed RG transformation successively generates the low-energy effective models describing the most relevant degrees of freedom with  preserving the criticality as fixed points, ensuring the invariance of CHS, and greatly simplifying the calculation of CHS by ruling out the irrelevant operators. These results pave the way for developments of advanced  analytical and numerical RG techniques on interacting non-Hermitian systems.

\emph{Coarse-graining of momentum space.--}--We demonstrate different RG schemes using the non-Hermitian SSH models with extended hoppings as typical examples. The general Hamiltonian we consider reads as,
\begin{equation}\label{eq1}
 H=\sum_{n,j,\mu\nu} t_{j,\mu\nu} (c^{\dagger}_{n+j,\mu}c_{n,\nu}+h.c.)+(\frac{\gamma_{j,\mu\nu}}{2}c^{\dagger}_{n+j,\mu}c_{n,\nu}-h.c.),
\end{equation}
where $c_{n,\mu}$ is the annihilation operator at site $n$ with sublattice $\mu=A,B$.  $t_{j,\mu\nu}$ represents for the first several nearest neighbor hoppings and $\gamma_{j,\mu\nu}$ indicates the non-Hermicity.  We consider in this work  up to the next nearest neighbor hopping terms as shown by Fig.2, where the sublattice indices are rearranged giving rise to parameters, $t_{i}$ and $\gamma_i$ with $i=1,...,4$. Models with different $t_i$ and $\gamma_i$ are considered in the following in order to exemplify the generality of our methods.	

%We discuss in this work different choices of parameters, $t_j$ and $\gamma_j$, which
%Similar to classical Ginzburg-Landau-type field theories, the standard renormalization group approach to fermionic states  is constructed from the coarse-graining in reciprocal space, where the fast modes with higher momentum are decimated.
\begin{figure}[thb]
\includegraphics[width=\linewidth]{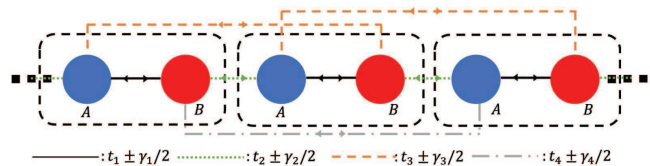}\label{fig2}
\caption{(color online). The SSH-type model with hopping $t_1$, $t_2$, $t_3$, $t_4$ and their non-Hermitian terms $\gamma_1$, $\gamma_2$, $\gamma_3$, $\gamma_4$. $t_1=t_{0,AB},\gamma_1=\gamma_{0,AB},t_2=t_{1,BA},\gamma_2=\gamma_{1,BA},t_3=t_{1,AB},\gamma_3=\gamma_{1,AB}$.}
\end{figure}

We firstly apply the momentum space RG to Eq.\eqref{eq1} \cite{sup} for a simple case with only nonzero $t_1$, $t_2$ and $\gamma_1$.
% We can formulate the functional action corresponding to Eq.\eqref{eq1} and make transformation to the lattice momentum space
The Hermitian SSH model with $\gamma_1=0$ is well studied; for $t_1\neq t_2$, the system acquires a mass gap  at $k=\pi$ (mod by 2$\pi$). Thus, one can obtain the long-distance continuum field theory near $k=\pi$, to which we perform the coarse-graining of fast modes, leading to the fixed point,  $t_1=t_2$. The system at  $t_1=t_2$ is gapless, scaling-invariant critical state between the topological trivial and nontrivial phase with winding number 0 and 1, respectively.
As expected, the standard method characterizes the critical state as fixed point in RG sense, as explicitly shown by the red line $t_1=t_2$ in Fig.2, which is the RG-invariant CHS embedded in  2D plane $(t_1,t_2)$.
\begin{figure}[thb]
\includegraphics[width=\linewidth]{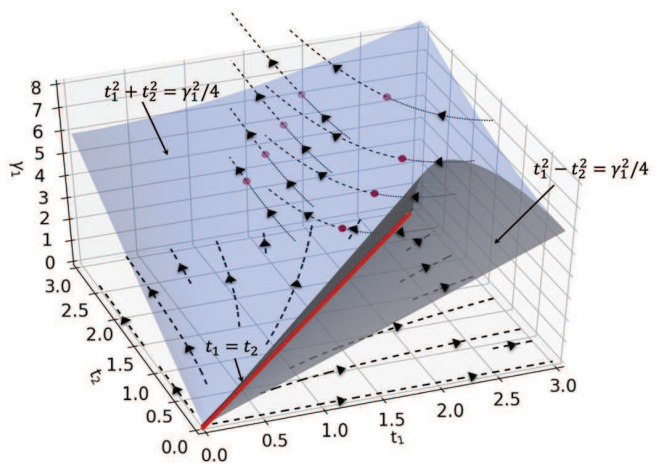}\label{fig3}
\caption{(color online).  RG flow obtained by momentum space coarse-graining, with nonzero bare parameters $t_1$, $t_2$ and $\gamma_1$. The $t_1$-$t_2$ plane corresponds to the Hermitian model, where a red straight line $t_1=t_2$ indicates the fixed point under RG. In the non-Hermitian region with nonzero $\gamma_1$, two CHSs  emerge as shown by the blue and grey surface. These are not invariant under the RG transformation. }
\end{figure}

After turning on $\gamma_1$, the CHS for the non-Hermtian model can be calculated utilizing the notion of the generalized Brillion zone (GBZ) \cite{article16,article33,article19,article34,article35}, leading to  $|t_1^2 - \frac{\gamma_1^2}{4}| = |t_2^2 - \frac{\gamma_2^2}{4}|$ \cite{sup}. As shown in Fig.3, two branches of CHS embedded in space $(t_1,t_2,\gamma_1)$ occur, where the grey surface intersects with the ``Hermitian plane" at  $t_1=t_2$ and with the blue surface at $t_1=\gamma_1/2$.
Then, we apply the same RG transformation as above to the long-distance continuum model derived near the blue branch of CHS
  \cite{sup}. It turns out that the CHS does not remain invariant as long as $\gamma_1\neq0$, as shown in Fig.3, the RG trajectories, i.e., the dashed curves with arrows, generically pass through the CHS, crossing from the trivial to nontrivial phase. The example shows that the non-Hermitian system at criticality is not respected by the standard momentum space RG.

%\begin{widetext}
%\begin{eqnarray}
%H=\sum_{k \in BZ}\begin{pmatrix}
%A^{\dagger}_k & B^{\dagger}_k
%\end{pmatrix}
%\begin{pmatrix}
%0  & ( t_1 + \frac{\gamma_1}{2} )+ (t_2 - \frac{\gamma_2}{2} ) e^{-ik} + (t_3 + \frac{\gamma_3}{2}) e^{ik} \\
%(t_1-\frac{\gamma_1}{2}) +(t_2 + \frac{\gamma_2}{2}) e^{ik} + (t_3-\frac{\gamma_3}{2} )e^{-ik}  & 0
%\end{pmatrix}
%\begin{pmatrix}  A_k \\ B_k
%\end{pmatrix}
%\end{eqnarray}
%\end{widetext}

\emph{Block-decimation in real space.--} The issue of the above conventional method is twofold. First, the decimation of the large momentum modes does not necessarily trace out the high energy windows with the proper ratio $b_r/b_i$, as indicated by Fig.1(d),  and second, it is known that the wave vector emergent from periodic boundary condition is not able to characterize intrinsic feature of non-Hermitian states.  We now propose a general block-decimation method to overcome these issues. 

Firstly, we rewrite Eq.\eqref{eq1} more compactly as, $H=\sum_n \sum_{j=-N}^{N}\sum_{\mu,\nu}\tilde{t}_{j,\mu\nu} c^{\dagger}_{n+j,\mu}c_{n,\nu}$, where the hopping $\tilde{t}_{j,\mu,\nu}$ has absorbed the non-Hermitian terms. Then, we formally reshape the 1D chain into a ladder composed of an upper and a lower chain, each of which consists of two sublattices in a unit cell,  as shown by Fig.4(a). Meanwhile, the total Hamiltonian is separated into three parts $H=H_{u}+H_{d}+H_{ud}$, where $H_{u}$ and $H_{d}$ describes the upper and lower chain  respectively, and $H_{ud}$ the interchain hopping. The partition function of the total system is then cast into $\mathcal{Z}=\int \mathcal{D}\overline{\psi}_u\mathcal{D}\psi_u\mathcal{D}\overline{\psi}_{d}\mathcal{D}\psi_{d}e^{-(S_{u}+S_d+S_{ud})}$,  where
\begin{equation}\label{eq2}
S_{a}[\bar{c},c] =-\sum_{i\omega_n}\sum_{k\in BZ^{\prime}} \bar{\psi}_{k,i\omega_n,a}(i\omega_n - \mathfrak{f}^0_{k})\psi_{k,i\omega_n,a},
\end{equation}
with $a=u,d$. $\psi_{k,i\omega_n,a}=[c_{k,i\omega_n,A},c_{k,i\omega_n,B}]^\mathrm{T}$ is the Grassmann field. $\mathfrak{f}^0_{k}$ is the kinetic term, a matrix in sublattice space whose elements are $f^0_{k,\mu\nu} = \sum_{j=-l}^{l} \tilde{t}_{2j,\mu\nu} e^{-i(2jk)} $with $l$ being the largest integer less than or equal to $N/2$. Since  the reciprocal space is also modified under decimation, the sum of $k$ in Eq.\eqref{eq2}  is restricted within $BZ^{\prime}$. The action for the interchain coupling can be written as,
\begin{equation}\label{eq3}
S_{ud}= \sum_{i\omega_n} \sum_{k \in BZ'} (\overline{\psi}_{k,i\omega_n,d}\mathfrak{f}^{t}_{k}\psi_{k,i\omega_n,u}+\bar{\psi}_{k,i\omega_n,u}\mathfrak{f}^{t}_{k}\psi_{k,i\omega_n,d}),
\end{equation}
where the components of the matrix $\mathfrak{f}^{t}_{k}$ reads as $f^{t}_{k,\mu\nu} = \sum_{j=-N+l+1}^{N-l} \tilde{t}_{2j-1,\mu\nu}e^{-i(2j-1)k}$.
Since the action is bilinear in terms of Grassmann fields, we exactly integrate out the lower chain as indicated by Fig.4(a), generating an effective action for the renormalized upper chain as,
\begin{equation}\label{eq4}
S^{\prime}_{u} = - \sum_{i\omega_n} \sum_{k \in BZ'}  \bar{\psi}_{k,i\omega_n,u}G^{-1}(i\omega_n,k)\psi_{k,i\omega_n,u}.
\end{equation}
The retarded Green's function $G(\omega,k)=\omega^+ - \mathfrak{f}^0_{k} - \mathfrak{f}^t_{k} (\omega^+- \mathfrak{f}^0_k)^{-1} \mathfrak{f}^t_{k}$ with $\omega^+=\omega+i0^+$ is obtained from the  Matsubara Green's function by analytic continuation. Since our main focus in this work is to investigate the critical points which take place in the low-energy window, it is convenient  to take the low-frequency limit in $G(\omega,k)$, which well preserves the low-energy states as long as the lower chain remains gapped. Then, one can read off an effective Hamiltonian from Eq.\eqref{eq4} as,
\begin{equation}\label{eq5}
\begin{split}
H_{eff} = \sum_{k \in BZ^{\prime}} \psi^{\dagger}_{k,u} (\mathfrak{f}^0_{k}-\mathfrak{f}^t_{k}(\mathfrak{f}^0_{k})^{-1} \mathfrak{f}^t_{k}) \psi_{k,u}
\end{split}
\end{equation}
which, after rescaling back to the original BZ, serves as the starting point for the next RG step.
 %The schematic plot that illustrates the RG procedure  using a specific model is shown by Fig.3, and this method can be readily applied to a general 1D non-Hermitian model \cite{sup}.

The block-decimation was firstly invented for classical and quantum spin models \cite{article94,article96}. its application to the non-Hermitian topological systems in the fashion of functional representation has not been explored to date. Unlike the momentum  coarse-graining, which inevitably introduces bias between the decimation of the real and imaginary energy modes, the real-space approach treats them on equal footing because both are traced out in the same manner.

Now we apply the block-decimation RG to some specific models. The first example we consider has nonzero parameters, $t_i$ and $\gamma_i$ with $i=1,2,3$, as shown by Fig.2. Following the above steps, we obtin
%we separate the chain into $H_{u,d}$ with interchain hopping $H_{ud}$.
 %In the basis  $\psi_{k,a}=[c_{k,A,a},c_{k,B,a}]^{\mathrm{T}}$  where $a=u,d$, $H_0=H_u+H_d$ is cast into, $H_0=\sum_{k,a} \psi^{\dagger}_{k,a}(\sigma^1t_1+i\sigma^2\gamma_1/2)\psi_{k,a}$.
%We note that the dispersion for the single chain is flat and has a full gap, since it only contains isolated unit cells as
% The interchain hopping reads as $H_{ud}=\sum_{k,a}\psi^{\dagger}_{k,a}(\sigma^1((t_2+t_3)cosk+i(\gamma_2/2+\gamma_3/2)sink)+\sigma^2((t_2-t_3)sink+i(-\gamma_2/2+\gamma_3/2)cosk)\psi_{k,\bar{a}}$.
%After integrating out the lower chain, scaling, and then taking the low-frequency approximation,
$H_{eff}$ describing the renormalized upper chain, which shares the same form as the bare one but with renormalized parameters.
%after and rescaling the lattice to original coordinate $x^{\prime}=x/2 $ $(k^{\prime}=2k)$,
The procedure generates a relation between the bare parameters $\mathbf{g}=\{t_1,t_2,t_3,\gamma_1,\gamma_2,\gamma_3\}$ and the renormalized ones $\mathbf{g}^{\prime}=\{t^{\prime}_1,t^{\prime}_2,t^{\prime}_3,\gamma^{\prime}_1,\gamma^{\prime}_2,\gamma^{\prime}_3\}$ , i.e., $\mathbf{g}^{\prime}=\mathcal{R}[\mathbf{g}]$, which is a complicated algebraic mapping in the 6D parameter space \cite{sup}.

%Noting that, for SSH-type Hamiltonian with longer-range hoppings, it is not an easy task to accurately obtain the critical surface of the model in Fig3, which by itself is a high dimensional manifold embedded in $\mathbb{R}^6$. The currently known method relies on the knowledge of GBZ \cite{article16,article33,article19,article34,article35}, the difficulty to determine which grows exponentially with increasing number of hopping terms.

An analytic solution of the mapping $\mathcal{R}$ can be obtained for $t_3=\gamma_3=0$, which leads to $t^{\prime}_1=t_1$, $\gamma^{\prime}_1=\gamma_1$, $t'_2 = - [2t_2\gamma_1\gamma_2+t_1(4t^2_2+\gamma^2_2)]/(4t^2_1-\gamma^2_1)$ and $ \gamma'_2 = - (4t^2_2\gamma_1+8t_1t_2\gamma_2+\gamma_1\gamma^2_2)/(4t^2_1-\gamma^2_1)$. Using the above relations, one then obtains
$ t'^2_1 - \gamma'^2_1/4 =  t^2_1 - \gamma^2_1/4 $, $ t'^2_2 - \gamma'^2_2/4 = (t^2_2 - \gamma^2_2/4)^2/[t^2_1 - (\gamma^2_1/4)]$. On the other hand, the CHS in this case can be exactly found as a hypersuface in $\mathbb{R}^4$ satisfying  $|t^2_1-\gamma^2_1/4|=|t^2_2-\gamma^2_2/4|$ \cite{sup}. Given a bare model located in the CHS, it then becomes obvious that $|t^{\prime2}_1-\gamma^{\prime2}_1/4|=|t^{\prime2}_2-\gamma^{\prime2}_2/4|$ is always satisfied after any steps of the RG mapping $\mathcal{R}.$, clearly showing the invariance of the CHS. For more complicated case with $t_3\neq0$ and $\gamma_3\neq0$,  we numerically evaluate the winding number $W$ \cite{article16,article19} for each RG step $l$, whose sudden change indicates the critical points as a function of $l$.
% Then, through monitoring the critical points under RG iteration, one can observe whether the CHS is closed or not.
%Introducing the generalized Bloch phase factor $\beta=e^{ikr} (k \in \mathbb{C})$ ,  the model is then described by the generalized Bloch Hamiltonian in the GBZ as $H(\beta)=\sigma^+R_+(\beta)+\sigma^-R_-(\beta)$, where $\sigma^{\pm}=\sigma^1\pm i\sigma^2$, $R_{+}(\beta)=(t_1+\gamma_1/2)+(t_2-\gamma_2/2)\beta^{-1}+(t_3+\gamma_3/2)\beta$, and $R_{-}(\beta)=(t_1-\gamma_1/2)+(t_2+\gamma_2/2)\beta^{-1}+(t_3-\gamma_3/2)\beta^{-1}$. The winding number is then calculated as
%\begin{equation}\label{eq6}
%W = - \frac{1}{2\pi} \int_{{C_{\beta}}}\frac{\mathrm{arg} [R_+(\beta)]-\mathrm{arg}[R_-(\beta)]}{2}.
%\end{equation}
Fig.4(b) shows the winding numbers as a function $t_2$ with other parameters fixed. The sudden change of $W$ indicates the critical point being around $t_{2c}=1.5$. The blue and red horizontal data lines in the inset of Fig.4(b) show the positions of $t_{2c}$ as a function of the RG step $l$, for $t_1=1.8$ and $t_1=1.0$, respectively, clearly displaying the invariance of the critical points under the proposed RG transformation.

As another example, we consider a  slightly different  model with $t_3=0$ and $t_4\neq0$ in Fig.2. In this case, longer-range hopping terms absent in the bare Hamiltonian emerge after RG transformation \cite{sup}.  To testify whether the CHS is invariant, we calculate the energy spectrum $|E|$ for the renormalized model, $H_{eff}$, under open boundaries, where the critical points are indicated by the gapless nodes $|E|=0$. On the other hand, the critical points of the bare model can be accurately obtained by examining the winding number $W$. As shown by Fig. 4(c), the critical points of the renormalized and bare model are found to match exactly with each other.  Therefore, the block decimation still respects the CHS even if new hopping terms are generated under RG.  

More generally, suppose a new set of hopping terms $\mathbf{t}_{a}$ emerge that are absent in $\mathbf{g}$ of the bare model. In this case, a CHS described by $\mathcal{S}(\mathbf{g},\mathbf{t}_{a})=\mathcal{S}(\mathcal{R}(\mathbf{g}))=0$,  can in principle be found, but embedded in the higher dimensional space of renormalized parameters, $\mathbb{R}^{D}$, where $D=\mathrm{dim}[\mathbf{g}]+\mathrm{dim}[\mathbf{t}_a]$.
The ``intersection" between $\mathcal{S}(\mathbf{g},\mathbf{t}_{a})=0$ and the manifold $\mathbb{R}^{\mathrm{dim}[\mathbf{g}]}$ automatically gives rise to a submanifold of $\mathcal{S}(\mathbf{g},\mathbf{t}_{a})=0$, which is nothing but the CHS embedded in  $\mathbb{R}^{\mathrm{dim}[\mathbf{g}]}$, i.e., $\mathcal{S}_0(\mathbf{g})=0$. Being a submanifold, $\mathcal{S}_0(\mathbf{g})=0$ is thereby equivalent to the ``intersection" of the two CHSs, $\mathcal{S}(\mathbf{g},\mathbf{t}_a)=\mathcal{S}_0(\mathbf{g})=0$, therefore $\mathcal{S}(\mathcal{R}(\mathbf{g}))=\mathcal{S}_0(\mathbf{g})=0$, i.e., the CHS remains intact under the proposed block-decimation.
%Since $\mathcal{S}(\mathbf{g},\mathbf{t}_{a})=0$  and $\mathcal{S}_0(\mathbf{g})=0$, are connected simply via $\mathcal{R}$, i.e., $\mathcal{S}(\mathbf{g},\mathbf{t}_{a})=0$ being equivalent to $\mathcal{S}(\mathcal{R}(\mathbf{g}))=0$, moreover, the intesection $\mathcal{S}_0(\mathbf{g})=\mathcal{S}(\mathcal{R}(\mathbf{g}))=0$ is equivalent to $\mathcal{S}_0(\mathbf{g})=0$ since the latter is the submanifold of $\mathcal{S}(\mathbf{g},\mathbf{t}_{a})=0$,  ensuring that the CHS of $H_{eff}$ remains the same with that of the bare model.
\begin{figure}[thb]
\includegraphics[width=\linewidth]{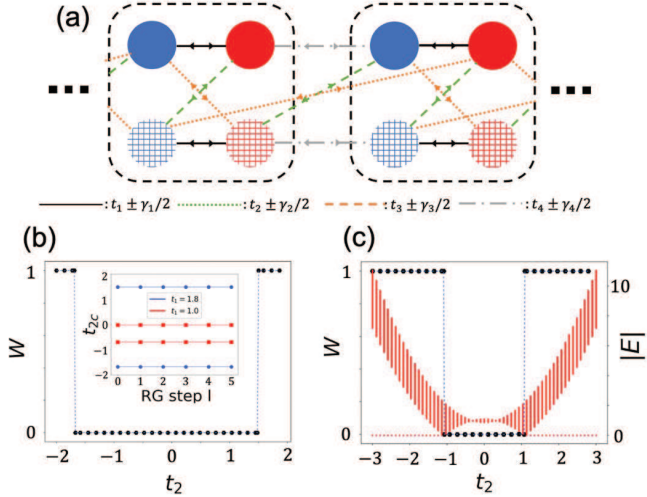}\label{fig4}
\caption{(color online). Block-decimation of SSH chain and the invariance of CHS. (a) Deformation of Fig(2) into a ladder, where the even unit cells are pull down to form the lower chain. The gridded sites in the lower-chain are integrated out. (b) Calculated winding number $W$ as a function of $t_2$, with $t_1 = 1.8$, $t_3 = 1/5$, $\gamma_1 = 5/3$, $\gamma_2 = 1/3$, $\gamma_3 = 1/6$. The inset shows that the critical points $t_{2c}$ do not change with RG step $l$.  (c) The spectrum of the effective Hamiltonian for $l=1$ and the winding number $W$ of the bare model.  $t_1=1$,  $t_4=0.1$, $\gamma_1=1/3$.  The gapless nodes of renormalized spectrum agrees exactly with the critical points of the bare model. This validates the invariance of CHS and also justifies the bulk-edge correspondence in the non-Bloch topological band theory. The values of  parameters are taken zero if not claimed. }
\end{figure}

\emph{An analytic proof.--}  We now provide a rigorous proof for the following claim, i.e., the block-decimation preserves the CHS of any 1D non-Hermitian systems as long as the subsystem to be traced out, e.g., the lower chain in Fig.2, remains gapped. To achieve so, we start from the general renormalized Hamiltonian $H_{eff}$ in Eq.\eqref{eq5} under the open boundaries. Then, we promote the Bloch wave-vector to $k=-iln\beta$ and modify BZ to the GBZ \cite{article19}. Since the CHS for the $H_{eff}$ at each RG step is determined by the gapless nodes in GBZ with $|E|=0$, it is sufficient to prove that the gapless nodes remain unchanged under the block-decimation.

Suppose that  $H^{(l)}_{eff}$ at the $l$-th RG step is obtained. Using Eq.\eqref{eq2} and Eq.\eqref{eq3}, it can be written into a matrix  as, $\mathcal{H}^{(l)}_{eff}=\tau^0\mathfrak{f}^{(l),0}_{\beta}+\tau^x\mathfrak{f}^{(l),t}_{\beta}$, where $\boldsymbol{\tau}$ denotes the upper/lower chain space and $\mathfrak{f}^{(l),0}_{\beta}$, $\mathfrak{f}^{(l),t}_{\beta}$ are the matrices at  the $l$-th step, similar to those of Eq.\eqref{eq5}.
%(in the Hermitian limit $\mathfrak{f}^{(l),0}_{\beta}=\mathfrak{f}^0_{-\frac{i}{2}ln\beta}$ ) 
Then for the $l+1$-th step, the GBZ can be uniquely determined  by the eigen-equation, $H^{(l+1)}_{eff}\alpha = E^{(l+1)} \alpha$, where $H^{(l+1)}_{eff}=\mathfrak{f}^{(l),0}_{\beta}-\mathfrak{f}^{(l),t}_{\beta}(\mathfrak{f}^{(l),0}_{\beta})^{-1} \mathfrak{f}^{(l),t}_{\beta} $
, as well as a condition $|\beta_{M}|=|\beta_{M+1}|$, where $\beta_i$ with $ {i=1,2,...,2M}$ are the solutions of the eigen-equation satisfying $|\beta_1|\le|\beta_2|\le...\le|\beta_{2M}|$. Here, $2M$ is the degree of the algebraic equation for $\beta$ \cite{article19}.  Moreover, as long as there exists gapless nodes in the spectrum of $H^{(l+1)}_{eff}$, they are determined by following equation,
\begin{equation}\label{eq7}
\begin{split}
\mathrm{det}[H^{(l+1)}_{eff}] = 0, ~\mathrm{and}~ \beta \in \mathrm{GBZ}.
\end{split}
\end{equation}
The first condition in Eq.\eqref{eq7} generates $\beta_i$ as function of  the model parameters $\mathbf{g}_l$ at step $l$, since both $\mathfrak{f}^{(l),0}_{\beta}$ and $\mathfrak{f}^{(l),t}_{\beta}$ are dependent on $\mathbf{g}_l$. The second condition $\beta\in\mathrm{GBZ}$ further requires that $|\beta_{M}(\mathbf{g}_l)|=|\beta_{M+1}(\mathbf{g}_l)|$, which uniquely produces the implicit expression of the CHS at the $l+1$-th step. Similar to Eq.\eqref{eq7}, the CHS at the $l$-th step is then described by
 \begin{equation}\label{eq8}
\mathrm{det}[\mathcal{H}^{(l)}_{eff}]=0, ~\mathrm{and}~ \beta \in \mathrm{GBZ}.
\end{equation}
Last, we note that Eq.\eqref{eq7} and Eq.\eqref{eq8} are related to each other via a simple matrix product as,
\begin{equation}
\begin{aligned}
& 
H^{(l)}_{eff}\begin{pmatrix}
1  &  0  \\
-(\mathfrak{f}^0_{\beta})^{-1} \mathfrak{f}^t_{\beta}  &  (\mathfrak{f}^0_{\beta})^{-1}
\end{pmatrix} =
\begin{pmatrix}
H^{(l+1)}_{eff} &  \mathfrak{f}^t_{\beta}(\mathfrak{f}^0_{\beta})^{-1}  \\
0 &  1
\end{pmatrix},
\end{aligned}
\end{equation}
such that we obtain  for $\beta\in\mathrm{GBZ}$ that
\begin{equation}	\label{eq9}
\mathrm{det}[H^{(l)}_{eff}]=0~~
\iff
\mathrm{det}[H^{(l+1)}_{eff}]= 0,
\end{equation}
as long as $\mathfrak{f}^0_{\beta}$ is invertible with  $\mathrm{det}[(\mathfrak{f}^0_{\beta})^{-1}] \neq 0$. Noting that this essentially requires that subsystem to be integrated out remains gapful, in consistent with what we assumed for the low-frequency approximation above and is satisfied by a large class of models as exemplified above. The equivalence in Eq.\eqref{eq9} proves that the effective Hamiltonian at two successive  RG step share the same equation of CHS, i.e., the claim raised above.

\emph{Conclusion and discussion.--} To summarize, we investigate the criticality of non-Hermitian system based on SSH-type models. We show that the standard RG with momentum coase-graining is inappropriate to non-Hermitian phases because it cannot preserve the criticality as RG fixed point. Instead, we propose a block-decimation scheme based on functional representation and prove that it produces RG flows that always respect the CHSs of non-Hermitian phases. It is worthwhile to point out that, the method can be easily applied to complicated models with longer-range hopping whose CHS is difficult to obtain \cite{sup}, and it turns out that the proposed RG method can rule out the irrelevant operators and greatly simplify the calculation of the CHS for the renormalized models \cite{sup}.  These results are not specific to SSH-type models but have general applications to other low-dimensional non-Hermitian quantum systems. The success of the real-space block-decimation sheds light onto further developments of numerical algorithms and perturbative RG calculations for the realm of non-Hermitian physics.

This work was supported by the National Program on Key Research Project (No. 2016YFA 0300501) and the Youth Program of National Natural Science Foundation of China (No. 11904225).


\begin{thebibliography}{68}%
\makeatletter
\providecommand \@ifxundefined [1]{%
 \@ifx{#1\undefined}
}%
\providecommand \@ifnum [1]{%
 \ifnum #1\expandafter \@firstoftwo
 \else \expandafter \@secondoftwo
 \fi
}%
\providecommand \@ifx [1]{%
 \ifx #1\expandafter \@firstoftwo
 \else \expandafter \@secondoftwo
 \fi
}%
\providecommand \natexlab [1]{#1}%
\providecommand \enquote  [1]{``#1''}%
\providecommand \bibnamefont  [1]{#1}%
\providecommand \bibfnamefont [1]{#1}%
\providecommand \citenamefont [1]{#1}%
\providecommand \href@noop [0]{\@secondoftwo}%
\providecommand \href [0]{\begingroup \@sanitize@url \@href}%
\providecommand \@href[1]{\@@startlink{#1}\@@href}%
\providecommand \@@href[1]{\endgroup#1\@@endlink}%
\providecommand \@sanitize@url [0]{\catcode `\\12\catcode `\$12\catcode
  `\&12\catcode `\#12\catcode `\^12\catcode `\_12\catcode `\%12\relax}%
\providecommand \@@startlink[1]{}%
\providecommand \@@endlink[0]{}%
\providecommand \url  [0]{\begingroup\@sanitize@url \@url }%
\providecommand \@url [1]{\endgroup\@href {#1}{\urlprefix }}%
\providecommand \urlprefix  [0]{URL }%
\providecommand \Eprint [0]{\href }%
\providecommand \doibase [0]{http://dx.doi.org/}%
\providecommand \selectlanguage [0]{\@gobble}%
\providecommand \bibinfo  [0]{\@secondoftwo}%
\providecommand \bibfield  [0]{\@secondoftwo}%
\providecommand \translation [1]{[#1]}%
\providecommand \BibitemOpen [0]{}%
\providecommand \bibitemStop [0]{}%
\providecommand \bibitemNoStop [0]{.\EOS\space}%
\providecommand \EOS [0]{\spacefactor3000\relax}%
\providecommand \BibitemShut  [1]{\csname bibitem#1\endcsname}%
\let\auto@bib@innerbib\@empty
%</preamble>
\bibitem [{\citenamefont {Kardar}(2007)}]{book1}%
  \BibitemOpen
  \bibfield  {author} {\bibinfo {author} {\bibfnamefont {Mehran}\ \bibnamefont
  {Kardar}},\ }\href {\doibase 10.1017/CBO9780511815881} {\emph {\bibinfo
  {title} {Statistical Physics of Fields}}}\ (\bibinfo {year}
  {2007})\BibitemShut {NoStop}%
\bibitem [{\citenamefont {Chen}\ \emph {et~al.}(2017)\citenamefont {Chen},
  \citenamefont {Legner}, \citenamefont {R\"uegg},\ and\ \citenamefont
  {Sigrist}}]{article2}%
  \BibitemOpen
  \bibfield  {author} {\bibinfo {author} {\bibfnamefont {Wei}\ \bibnamefont
  {Chen}}, \bibinfo {author} {\bibfnamefont {Markus}\ \bibnamefont {Legner}},
  \bibinfo {author} {\bibfnamefont {Andreas}\ \bibnamefont {R\"uegg}}, \ and\
  \bibinfo {author} {\bibfnamefont {Manfred}\ \bibnamefont {Sigrist}},\
  }\bibfield  {title} {\enquote {\bibinfo {title} {Correlation length,
  universality classes, and scaling laws associated with topological phase
  transitions},}\ }\href {\doibase 10.1103/PhysRevB.95.075116} {\bibfield
  {journal} {\bibinfo  {journal} {Phys. Rev. B}\ }\textbf {\bibinfo {volume}
  {95}},\ \bibinfo {pages} {075116} (\bibinfo {year} {2017})}\BibitemShut
  {NoStop}%
\bibitem [{\citenamefont {Chen}(2016)}]{article3}%
  \BibitemOpen
  \bibfield  {author} {\bibinfo {author} {\bibfnamefont {Wei}\ \bibnamefont
  {Chen}},\ }\bibfield  {title} {\enquote {\bibinfo {title} {Scaling theory of
  topological phase transitions},}\ }\href {\doibase
  10.1088/0953-8984/28/5/055601} {\bibfield  {journal} {\bibinfo  {journal} {J.
  Phys. Condens. Matter}\ }\textbf {\bibinfo {volume} {28}},\ \bibinfo {pages}
  {055601} (\bibinfo {year} {2016})}\BibitemShut {NoStop}%
\bibitem [{\citenamefont {Zhen}\ \emph {et~al.}(2015)\citenamefont {Zhen},
  \citenamefont {Hsu}, \citenamefont {Igarashi}, \citenamefont {Lu},
  \citenamefont {Kaminer}, \citenamefont {Pick}, \citenamefont {Chua},
  \citenamefont {Joannopoulos},\ and\ \citenamefont {Soljačić}}]{article4}%
  \BibitemOpen
  \bibfield  {author} {\bibinfo {author} {\bibfnamefont {Bo}~\bibnamefont
  {Zhen}}, \bibinfo {author} {\bibfnamefont {Chia~Wei}\ \bibnamefont {Hsu}},
  \bibinfo {author} {\bibfnamefont {Yuichi}\ \bibnamefont {Igarashi}}, \bibinfo
  {author} {\bibfnamefont {Ling}\ \bibnamefont {Lu}}, \bibinfo {author}
  {\bibfnamefont {Ido}\ \bibnamefont {Kaminer}}, \bibinfo {author}
  {\bibfnamefont {Adi}\ \bibnamefont {Pick}}, \bibinfo {author} {\bibfnamefont
  {Song-Liang}\ \bibnamefont {Chua}}, \bibinfo {author} {\bibfnamefont {John}\
  \bibnamefont {Joannopoulos}}, \ and\ \bibinfo {author} {\bibfnamefont
  {Marin}\ \bibnamefont {Soljačić}},\ }\bibfield  {title} {\enquote {\bibinfo
  {title} {Spawning rings of exceptional points out of dirac cones},}\ }\href
  {\doibase 10.1038/nature14889} {\bibfield  {journal} {\bibinfo  {journal}
  {Nature}\ }\textbf {\bibinfo {volume} {525}},\ \bibinfo {pages} {354--358}
  (\bibinfo {year} {2015})}\BibitemShut {NoStop}%
\bibitem [{\citenamefont {El-Ganainy}\ \emph {et~al.}(2018)\citenamefont
  {El-Ganainy}, \citenamefont {Makris}, \citenamefont {Khajavikhan},
  \citenamefont {Musslimani}, \citenamefont {Rotter},\ and\ \citenamefont
  {Christodoulides}}]{article5}%
  \BibitemOpen
  \bibfield  {author} {\bibinfo {author} {\bibfnamefont {Ramy}\ \bibnamefont
  {El-Ganainy}}, \bibinfo {author} {\bibfnamefont {Konstantinos}\ \bibnamefont
  {Makris}}, \bibinfo {author} {\bibfnamefont {Mercedeh}\ \bibnamefont
  {Khajavikhan}}, \bibinfo {author} {\bibfnamefont {Ziad}\ \bibnamefont
  {Musslimani}}, \bibinfo {author} {\bibfnamefont {Stefan}\ \bibnamefont
  {Rotter}}, \ and\ \bibinfo {author} {\bibfnamefont {Demetrios}\ \bibnamefont
  {Christodoulides}},\ }\bibfield  {title} {\enquote {\bibinfo {title}
  {Non-hermitian physics and pt symmetry},}\ }\href {\doibase
  10.1038/nphys4323} {\bibfield  {journal} {\bibinfo  {journal} {Nature
  Physics}\ }\textbf {\bibinfo {volume} {14}},\ \bibinfo {pages} {11--19}
  (\bibinfo {year} {2018})}\BibitemShut {NoStop}%
\bibitem [{\citenamefont {Ozawa}\ \emph {et~al.}(2019)\citenamefont {Ozawa},
  \citenamefont {Price}, \citenamefont {Amo}, \citenamefont {Goldman},
  \citenamefont {Hafezi}, \citenamefont {Lu}, \citenamefont {Rechtsman},
  \citenamefont {Schuster}, \citenamefont {Simon}, \citenamefont {Zilberberg},\
  and\ \citenamefont {Carusotto}}]{article6}%
  \BibitemOpen
  \bibfield  {author} {\bibinfo {author} {\bibfnamefont {Tomoki}\ \bibnamefont
  {Ozawa}}, \bibinfo {author} {\bibfnamefont {Hannah~M.}\ \bibnamefont
  {Price}}, \bibinfo {author} {\bibfnamefont {Alberto}\ \bibnamefont {Amo}},
  \bibinfo {author} {\bibfnamefont {Nathan}\ \bibnamefont {Goldman}}, \bibinfo
  {author} {\bibfnamefont {Mohammad}\ \bibnamefont {Hafezi}}, \bibinfo {author}
  {\bibfnamefont {Ling}\ \bibnamefont {Lu}}, \bibinfo {author} {\bibfnamefont
  {Mikael~C.}\ \bibnamefont {Rechtsman}}, \bibinfo {author} {\bibfnamefont
  {David}\ \bibnamefont {Schuster}}, \bibinfo {author} {\bibfnamefont
  {Jonathan}\ \bibnamefont {Simon}}, \bibinfo {author} {\bibfnamefont {Oded}\
  \bibnamefont {Zilberberg}}, \ and\ \bibinfo {author} {\bibfnamefont {Iacopo}\
  \bibnamefont {Carusotto}},\ }\bibfield  {title} {\enquote {\bibinfo {title}
  {Topological photonics},}\ }\href {\doibase 10.1103/RevModPhys.91.015006}
  {\bibfield  {journal} {\bibinfo  {journal} {Rev. Mod. Phys.}\ }\textbf
  {\bibinfo {volume} {91}},\ \bibinfo {pages} {015006} (\bibinfo {year}
  {2019})}\BibitemShut {NoStop}%
\bibitem [{\citenamefont {Feng}\ \emph {et~al.}(2017)\citenamefont {Feng},
  \citenamefont {El-Ganainy},\ and\ \citenamefont {Ge}}]{article7}%
  \BibitemOpen
  \bibfield  {author} {\bibinfo {author} {\bibfnamefont {Liang}\ \bibnamefont
  {Feng}}, \bibinfo {author} {\bibfnamefont {Ramy}\ \bibnamefont {El-Ganainy}},
  \ and\ \bibinfo {author} {\bibfnamefont {Li}~\bibnamefont {Ge}},\ }\bibfield
  {title} {\enquote {\bibinfo {title} {Non-hermitian photonics based on
  parity–time symmetry},}\ }\href {https://doi.org/10.1038/s41566-017-0031-1}
  {\bibfield  {journal} {\bibinfo  {journal} {Nat. Photonics}\ }\textbf
  {\bibinfo {volume} {11}} (\bibinfo {year} {2017})}\BibitemShut {NoStop}%
\bibitem [{\citenamefont {Zhou}\ \emph {et~al.}(2018)\citenamefont {Zhou},
  \citenamefont {Peng}, \citenamefont {Yoon}, \citenamefont {Hsu},
  \citenamefont {Nelson}, \citenamefont {Fu}, \citenamefont {Joannopoulos},
  \citenamefont {Solja{\v c}i{\'c}},\ and\ \citenamefont {Zhen}}]{article8}%
  \BibitemOpen
  \bibfield  {author} {\bibinfo {author} {\bibfnamefont {Hengyun}\ \bibnamefont
  {Zhou}}, \bibinfo {author} {\bibfnamefont {Chao}\ \bibnamefont {Peng}},
  \bibinfo {author} {\bibfnamefont {Yoseob}\ \bibnamefont {Yoon}}, \bibinfo
  {author} {\bibfnamefont {Chia~Wei}\ \bibnamefont {Hsu}}, \bibinfo {author}
  {\bibfnamefont {Keith~A.}\ \bibnamefont {Nelson}}, \bibinfo {author}
  {\bibfnamefont {Liang}\ \bibnamefont {Fu}}, \bibinfo {author} {\bibfnamefont
  {John~D.}\ \bibnamefont {Joannopoulos}}, \bibinfo {author} {\bibfnamefont
  {Marin}\ \bibnamefont {Solja{\v c}i{\'c}}}, \ and\ \bibinfo {author}
  {\bibfnamefont {Bo}~\bibnamefont {Zhen}},\ }\bibfield  {title} {\enquote
  {\bibinfo {title} {Observation of bulk fermi arc and polarization half charge
  from paired exceptional points},}\ }\href {\doibase 10.1126/science.aap9859}
  {\bibfield  {journal} {\bibinfo  {journal} {Science}\ }\textbf {\bibinfo
  {volume} {359}},\ \bibinfo {pages} {1009--1012} (\bibinfo {year}
  {2018})}\BibitemShut {NoStop}%
\bibitem [{\citenamefont {Kozii}\ and\ \citenamefont {Fu}()}]{article9}%
  \BibitemOpen
  \bibfield  {author} {\bibinfo {author} {\bibfnamefont {Vladyslav}\
  \bibnamefont {Kozii}}\ and\ \bibinfo {author} {\bibfnamefont {Liang}\
  \bibnamefont {Fu}},\ }\href {https://arxiv.org/abs/1708.05841} {\enquote
  {\bibinfo {title} {Non-hermitian topological theory of finite-lifetime
  quasiparticles: Prediction of bulk fermi arc due to exceptional point},}\
  }\BibitemShut {NoStop}%
\bibitem [{\citenamefont {Yoshida}\ \emph {et~al.}(2018)\citenamefont
  {Yoshida}, \citenamefont {Peters},\ and\ \citenamefont
  {Kawakami}}]{article10}%
  \BibitemOpen
  \bibfield  {author} {\bibinfo {author} {\bibfnamefont {Tsuneya}\ \bibnamefont
  {Yoshida}}, \bibinfo {author} {\bibfnamefont {Robert}\ \bibnamefont
  {Peters}}, \ and\ \bibinfo {author} {\bibfnamefont {Norio}\ \bibnamefont
  {Kawakami}},\ }\bibfield  {title} {\enquote {\bibinfo {title} {Non-hermitian
  perspective of the band structure in heavy-fermion systems},}\ }\href
  {\doibase 10.1103/PhysRevB.98.035141} {\bibfield  {journal} {\bibinfo
  {journal} {Phys. Rev. B}\ }\textbf {\bibinfo {volume} {98}},\ \bibinfo
  {pages} {035141} (\bibinfo {year} {2018})}\BibitemShut {NoStop}%
\bibitem [{\citenamefont {Papaj}\ \emph {et~al.}(2019)\citenamefont {Papaj},
  \citenamefont {Isobe},\ and\ \citenamefont {Fu}}]{article11}%
  \BibitemOpen
  \bibfield  {author} {\bibinfo {author} {\bibfnamefont {Micha\l{}}\
  \bibnamefont {Papaj}}, \bibinfo {author} {\bibfnamefont {Hiroki}\
  \bibnamefont {Isobe}}, \ and\ \bibinfo {author} {\bibfnamefont {Liang}\
  \bibnamefont {Fu}},\ }\bibfield  {title} {\enquote {\bibinfo {title} {Nodal
  arc of disordered dirac fermions and non-hermitian band theory},}\ }\href
  {\doibase 10.1103/PhysRevB.99.201107} {\bibfield  {journal} {\bibinfo
  {journal} {Phys. Rev. B}\ }\textbf {\bibinfo {volume} {99}},\ \bibinfo
  {pages} {201107} (\bibinfo {year} {2019})}\BibitemShut {NoStop}%
\bibitem [{\citenamefont {McClarty}\ and\ \citenamefont
  {Rau}(2019)}]{article12}%
  \BibitemOpen
  \bibfield  {author} {\bibinfo {author} {\bibfnamefont {Paul~A.}\ \bibnamefont
  {McClarty}}\ and\ \bibinfo {author} {\bibfnamefont {Jeffrey~G.}\ \bibnamefont
  {Rau}},\ }\bibfield  {title} {\enquote {\bibinfo {title} {Non-hermitian
  topology of spontaneous magnon decay},}\ }\href {\doibase
  10.1103/PhysRevB.100.100405} {\bibfield  {journal} {\bibinfo  {journal}
  {Phys. Rev. B}\ }\textbf {\bibinfo {volume} {100}},\ \bibinfo {pages}
  {100405} (\bibinfo {year} {2019})}\BibitemShut {NoStop}%
\bibitem [{\citenamefont {Shen}\ and\ \citenamefont
  {Fu}(2018{\natexlab{a}})}]{article13}%
  \BibitemOpen
  \bibfield  {author} {\bibinfo {author} {\bibfnamefont {Huitao}\ \bibnamefont
  {Shen}}\ and\ \bibinfo {author} {\bibfnamefont {Liang}\ \bibnamefont {Fu}},\
  }\bibfield  {title} {\enquote {\bibinfo {title} {Quantum oscillation from
  in-gap states and a non-hermitian landau level problem},}\ }\href {\doibase
  10.1103/PhysRevLett.121.026403} {\bibfield  {journal} {\bibinfo  {journal}
  {Phys. Rev. Lett.}\ }\textbf {\bibinfo {volume} {121}},\ \bibinfo {pages}
  {026403} (\bibinfo {year} {2018}{\natexlab{a}})}\BibitemShut {NoStop}%
\bibitem [{\citenamefont {Shen}\ and\ \citenamefont
  {Fu}(2018{\natexlab{b}})}]{article14}%
  \BibitemOpen
  \bibfield  {author} {\bibinfo {author} {\bibfnamefont {Huitao}\ \bibnamefont
  {Shen}}\ and\ \bibinfo {author} {\bibfnamefont {Liang}\ \bibnamefont {Fu}},\
  }\bibfield  {title} {\enquote {\bibinfo {title} {Quantum oscillation from
  in-gap states and a non-hermitian landau level problem},}\ }\href {\doibase
  10.1103/PhysRevLett.121.026403} {\bibfield  {journal} {\bibinfo  {journal}
  {Phys. Rev. Lett.}\ }\textbf {\bibinfo {volume} {121}},\ \bibinfo {pages}
  {026403} (\bibinfo {year} {2018}{\natexlab{b}})}\BibitemShut {NoStop}%
\bibitem [{\citenamefont {Lee}(2016)}]{article15}%
  \BibitemOpen
  \bibfield  {author} {\bibinfo {author} {\bibfnamefont {Tony~E.}\ \bibnamefont
  {Lee}},\ }\bibfield  {title} {\enquote {\bibinfo {title} {Anomalous edge
  state in a non-hermitian lattice},}\ }\href {\doibase
  10.1103/PhysRevLett.116.133903} {\bibfield  {journal} {\bibinfo  {journal}
  {Phys. Rev. Lett.}\ }\textbf {\bibinfo {volume} {116}},\ \bibinfo {pages}
  {133903} (\bibinfo {year} {2016})}\BibitemShut {NoStop}%
\bibitem [{\citenamefont {Yao}\ and\ \citenamefont {Wang}(2018)}]{article16}%
  \BibitemOpen
  \bibfield  {author} {\bibinfo {author} {\bibfnamefont {Shunyu}\ \bibnamefont
  {Yao}}\ and\ \bibinfo {author} {\bibfnamefont {Zhong}\ \bibnamefont {Wang}},\
  }\bibfield  {title} {\enquote {\bibinfo {title} {Edge states and topological
  invariants of non-hermitian systems},}\ }\href {\doibase
  10.1103/PhysRevLett.121.086803} {\bibfield  {journal} {\bibinfo  {journal}
  {Phys. Rev. Lett.}\ }\textbf {\bibinfo {volume} {121}},\ \bibinfo {pages}
  {086803} (\bibinfo {year} {2018})}\BibitemShut {NoStop}%
\bibitem [{\citenamefont {Leykam}\ \emph {et~al.}(2017)\citenamefont {Leykam},
  \citenamefont {Bliokh}, \citenamefont {Huang}, \citenamefont {Chong},\ and\
  \citenamefont {Nori}}]{article17}%
  \BibitemOpen
  \bibfield  {author} {\bibinfo {author} {\bibfnamefont {Daniel}\ \bibnamefont
  {Leykam}}, \bibinfo {author} {\bibfnamefont {Konstantin~Y.}\ \bibnamefont
  {Bliokh}}, \bibinfo {author} {\bibfnamefont {Chunli}\ \bibnamefont {Huang}},
  \bibinfo {author} {\bibfnamefont {Y.~D.}\ \bibnamefont {Chong}}, \ and\
  \bibinfo {author} {\bibfnamefont {Franco}\ \bibnamefont {Nori}},\ }\bibfield
  {title} {\enquote {\bibinfo {title} {Edge modes, degeneracies, and
  topological numbers in non-hermitian systems},}\ }\href {\doibase
  10.1103/PhysRevLett.118.040401} {\bibfield  {journal} {\bibinfo  {journal}
  {Phys. Rev. Lett.}\ }\textbf {\bibinfo {volume} {118}},\ \bibinfo {pages}
  {040401} (\bibinfo {year} {2017})}\BibitemShut {NoStop}%
\bibitem [{\citenamefont {Kunst}\ \emph {et~al.}(2018)\citenamefont {Kunst},
  \citenamefont {Edvardsson}, \citenamefont {Budich},\ and\ \citenamefont
  {Bergholtz}}]{article18}%
  \BibitemOpen
  \bibfield  {author} {\bibinfo {author} {\bibfnamefont {Flore~K.}\
  \bibnamefont {Kunst}}, \bibinfo {author} {\bibfnamefont {Elisabet}\
  \bibnamefont {Edvardsson}}, \bibinfo {author} {\bibfnamefont {Jan~Carl}\
  \bibnamefont {Budich}}, \ and\ \bibinfo {author} {\bibfnamefont {Emil~J.}\
  \bibnamefont {Bergholtz}},\ }\bibfield  {title} {\enquote {\bibinfo {title}
  {Biorthogonal bulk-boundary correspondence in non-hermitian systems},}\
  }\href {\doibase 10.1103/PhysRevLett.121.026808} {\bibfield  {journal}
  {\bibinfo  {journal} {Phys. Rev. Lett.}\ }\textbf {\bibinfo {volume} {121}},\
  \bibinfo {pages} {026808} (\bibinfo {year} {2018})}\BibitemShut {NoStop}%
\bibitem [{\citenamefont {Yokomizo}\ and\ \citenamefont
  {Murakami}(2019)}]{article19}%
  \BibitemOpen
  \bibfield  {author} {\bibinfo {author} {\bibfnamefont {Kazuki}\ \bibnamefont
  {Yokomizo}}\ and\ \bibinfo {author} {\bibfnamefont {Shuichi}\ \bibnamefont
  {Murakami}},\ }\bibfield  {title} {\enquote {\bibinfo {title} {Non-bloch band
  theory of non-hermitian systems},}\ }\href {\doibase
  10.1103/PhysRevLett.123.066404} {\bibfield  {journal} {\bibinfo  {journal}
  {Phys. Rev. Lett.}\ }\textbf {\bibinfo {volume} {123}},\ \bibinfo {pages}
  {066404} (\bibinfo {year} {2019})}\BibitemShut {NoStop}%
\bibitem [{\citenamefont {Martinez~Alvarez}\ \emph
  {et~al.}(2018{\natexlab{a}})\citenamefont {Martinez~Alvarez}, \citenamefont
  {Barrios~Vargas}, \citenamefont {Berdakin},\ and\ \citenamefont
  {Foa~Torres}}]{article20}%
  \BibitemOpen
  \bibfield  {author} {\bibinfo {author} {\bibfnamefont {Victor}\ \bibnamefont
  {Martinez~Alvarez}}, \bibinfo {author} {\bibfnamefont {José~Eduardo}\
  \bibnamefont {Barrios~Vargas}}, \bibinfo {author} {\bibfnamefont {Matias}\
  \bibnamefont {Berdakin}}, \ and\ \bibinfo {author} {\bibfnamefont {Luis}\
  \bibnamefont {Foa~Torres}},\ }\bibfield  {title} {\enquote {\bibinfo {title}
  {Topological states of non-hermitian systems},}\ }\href
  {https://doi.org/10.1140/epjst/e2018-800091-5} {\bibfield  {journal}
  {\bibinfo  {journal} {Eur. Phys. J. Spec. Top.}\ }\textbf {\bibinfo {volume}
  {227}},\ \bibinfo {pages} {1295--1308} (\bibinfo {year}
  {2018}{\natexlab{a}})}\BibitemShut {NoStop}%
\bibitem [{\citenamefont {Zirnstein}\ \emph {et~al.}(2019)\citenamefont
  {Zirnstein}, \citenamefont {Refael},\ and\ \citenamefont
  {Rosenow}}]{article21}%
  \BibitemOpen
  \bibfield  {author} {\bibinfo {author} {\bibfnamefont {Heinrich-Gregor}\
  \bibnamefont {Zirnstein}}, \bibinfo {author} {\bibfnamefont {Gil}\
  \bibnamefont {Refael}}, \ and\ \bibinfo {author} {\bibfnamefont {Bernd}\
  \bibnamefont {Rosenow}},\ }\href {https://arxiv.org/abs/1901.11241} {\enquote
  {\bibinfo {title} {Bulk-boundary correspondence for non-hermitian
  hamiltonians via green functions},}\ } (\bibinfo {year} {2019})\BibitemShut
  {NoStop}%
\bibitem [{\citenamefont {Xiong}(2018)}]{article22}%
  \BibitemOpen
  \bibfield  {author} {\bibinfo {author} {\bibfnamefont {Ye}~\bibnamefont
  {Xiong}},\ }\bibfield  {title} {\enquote {\bibinfo {title} {Why does bulk
  boundary correspondence fail in some non-hermitian topological models},}\
  }\href {\doibase 10.1088/2399-6528/aab64a} {\bibfield  {journal} {\bibinfo
  {journal} {J. Phys. Commun.}\ }\textbf {\bibinfo {volume} {2}},\ \bibinfo
  {pages} {035043} (\bibinfo {year} {2018})}\BibitemShut {NoStop}%
\bibitem [{\citenamefont {Martinez~Alvarez}\ \emph
  {et~al.}(2018{\natexlab{b}})\citenamefont {Martinez~Alvarez}, \citenamefont
  {Barrios~Vargas},\ and\ \citenamefont {Foa~Torres}}]{article23}%
  \BibitemOpen
  \bibfield  {author} {\bibinfo {author} {\bibfnamefont {V.~M.}\ \bibnamefont
  {Martinez~Alvarez}}, \bibinfo {author} {\bibfnamefont {J.~E.}\ \bibnamefont
  {Barrios~Vargas}}, \ and\ \bibinfo {author} {\bibfnamefont {L.~E.~F.}\
  \bibnamefont {Foa~Torres}},\ }\bibfield  {title} {\enquote {\bibinfo {title}
  {Non-hermitian robust edge states in one dimension: Anomalous localization
  and eigenspace condensation at exceptional points},}\ }\href {\doibase
  10.1103/PhysRevB.97.121401} {\bibfield  {journal} {\bibinfo  {journal} {Phys.
  Rev. B}\ }\textbf {\bibinfo {volume} {97}},\ \bibinfo {pages} {121401}
  (\bibinfo {year} {2018}{\natexlab{b}})}\BibitemShut {NoStop}%
\bibitem [{\citenamefont {Lee}\ and\ \citenamefont
  {Thomale}(2019)}]{article24}%
  \BibitemOpen
  \bibfield  {author} {\bibinfo {author} {\bibfnamefont {Ching~Hua}\
  \bibnamefont {Lee}}\ and\ \bibinfo {author} {\bibfnamefont {Ronny}\
  \bibnamefont {Thomale}},\ }\bibfield  {title} {\enquote {\bibinfo {title}
  {Anatomy of skin modes and topology in non-hermitian systems},}\ }\href
  {\doibase 10.1103/PhysRevB.99.201103} {\bibfield  {journal} {\bibinfo
  {journal} {Phys. Rev. B}\ }\textbf {\bibinfo {volume} {99}},\ \bibinfo
  {pages} {201103} (\bibinfo {year} {2019})}\BibitemShut {NoStop}%
\bibitem [{\citenamefont {Jin}\ and\ \citenamefont {Song}(2019)}]{article25}%
  \BibitemOpen
  \bibfield  {author} {\bibinfo {author} {\bibfnamefont {L.}~\bibnamefont
  {Jin}}\ and\ \bibinfo {author} {\bibfnamefont {Z.}~\bibnamefont {Song}},\
  }\bibfield  {title} {\enquote {\bibinfo {title} {Bulk-boundary correspondence
  in a non-hermitian system in one dimension with chiral inversion symmetry},}\
  }\href {\doibase 10.1103/PhysRevB.99.081103} {\bibfield  {journal} {\bibinfo
  {journal} {Phys. Rev. B}\ }\textbf {\bibinfo {volume} {99}},\ \bibinfo
  {pages} {081103} (\bibinfo {year} {2019})}\BibitemShut {NoStop}%
\bibitem [{\citenamefont {Herviou}\ \emph {et~al.}(2019)\citenamefont
  {Herviou}, \citenamefont {Bardarson},\ and\ \citenamefont
  {Regnault}}]{article27}%
  \BibitemOpen
  \bibfield  {author} {\bibinfo {author} {\bibfnamefont {Lo\"{\i}c}\
  \bibnamefont {Herviou}}, \bibinfo {author} {\bibfnamefont {Jens~H.}\
  \bibnamefont {Bardarson}}, \ and\ \bibinfo {author} {\bibfnamefont {Nicolas}\
  \bibnamefont {Regnault}},\ }\bibfield  {title} {\enquote {\bibinfo {title}
  {Defining a bulk-edge correspondence for non-hermitian hamiltonians via
  singular-value decomposition},}\ }\href {\doibase 10.1103/PhysRevA.99.052118}
  {\bibfield  {journal} {\bibinfo  {journal} {Phys. Rev. A}\ }\textbf {\bibinfo
  {volume} {99}},\ \bibinfo {pages} {052118} (\bibinfo {year}
  {2019})}\BibitemShut {NoStop}%
\bibitem [{\citenamefont {Pocock}\ \emph {et~al.}(2019)\citenamefont {Pocock},
  \citenamefont {Huidobro},\ and\ \citenamefont {Giannini}}]{article28}%
  \BibitemOpen
  \bibfield  {author} {\bibinfo {author} {\bibfnamefont {Simon}\ \bibnamefont
  {Pocock}}, \bibinfo {author} {\bibfnamefont {Paloma}\ \bibnamefont
  {Huidobro}}, \ and\ \bibinfo {author} {\bibfnamefont {Vincenzo}\ \bibnamefont
  {Giannini}},\ }\bibfield  {title} {\enquote {\bibinfo {title} {Bulk-edge
  correspondence and long-range hopping in the topological plasmonic chain},}\
  }\href {https://doi.org/10.1515/nanoph-2019-0033} {\bibfield  {journal}
  {\bibinfo  {journal} {Nanophotonics}\ }\textbf {\bibinfo {volume} {8}},\
  \bibinfo {pages} {1337--1347} (\bibinfo {year} {2019})}\BibitemShut {NoStop}%
\bibitem [{\citenamefont {Wang}\ \emph {et~al.}(2019)\citenamefont {Wang},
  \citenamefont {Ruan},\ and\ \citenamefont {Zhang}}]{article66}%
  \BibitemOpen
  \bibfield  {author} {\bibinfo {author} {\bibfnamefont {Huaiqiang}\
  \bibnamefont {Wang}}, \bibinfo {author} {\bibfnamefont {Jiawei}\ \bibnamefont
  {Ruan}}, \ and\ \bibinfo {author} {\bibfnamefont {Haijun}\ \bibnamefont
  {Zhang}},\ }\bibfield  {title} {\enquote {\bibinfo {title} {Non-hermitian
  nodal-line semimetals with an anomalous bulk-boundary correspondence},}\
  }\href {\doibase 10.1103/PhysRevB.99.075130} {\bibfield  {journal} {\bibinfo
  {journal} {Phys. Rev. B}\ }\textbf {\bibinfo {volume} {99}},\ \bibinfo
  {pages} {075130} (\bibinfo {year} {2019})}\BibitemShut {NoStop}%
\bibitem [{\citenamefont {Yao}\ \emph {et~al.}(2018)\citenamefont {Yao},
  \citenamefont {Song},\ and\ \citenamefont {Wang}}]{article33}%
  \BibitemOpen
  \bibfield  {author} {\bibinfo {author} {\bibfnamefont {Shunyu}\ \bibnamefont
  {Yao}}, \bibinfo {author} {\bibfnamefont {Fei}\ \bibnamefont {Song}}, \ and\
  \bibinfo {author} {\bibfnamefont {Zhong}\ \bibnamefont {Wang}},\ }\bibfield
  {title} {\enquote {\bibinfo {title} {Non-hermitian chern bands},}\ }\href
  {\doibase 10.1103/PhysRevLett.121.136802} {\bibfield  {journal} {\bibinfo
  {journal} {Phys. Rev. Lett.}\ }\textbf {\bibinfo {volume} {121}},\ \bibinfo
  {pages} {136802} (\bibinfo {year} {2018})}\BibitemShut {NoStop}%
\bibitem [{\citenamefont {Deng}\ and\ \citenamefont {Yi}(2019)}]{article34}%
  \BibitemOpen
  \bibfield  {author} {\bibinfo {author} {\bibfnamefont {Tian-Shu}\
  \bibnamefont {Deng}}\ and\ \bibinfo {author} {\bibfnamefont {Wei}\
  \bibnamefont {Yi}},\ }\bibfield  {title} {\enquote {\bibinfo {title}
  {Non-bloch topological invariants in a non-hermitian domain wall system},}\
  }\href {\doibase 10.1103/PhysRevB.100.035102} {\bibfield  {journal} {\bibinfo
   {journal} {Phys. Rev. B}\ }\textbf {\bibinfo {volume} {100}},\ \bibinfo
  {pages} {035102} (\bibinfo {year} {2019})}\BibitemShut {NoStop}%
\bibitem [{\citenamefont {Liu}\ \emph {et~al.}(2019)\citenamefont {Liu},
  \citenamefont {Zhang}, \citenamefont {Ai}, \citenamefont {Gong},
  \citenamefont {Kawabata}, \citenamefont {Ueda},\ and\ \citenamefont
  {Nori}}]{article35}%
  \BibitemOpen
  \bibfield  {author} {\bibinfo {author} {\bibfnamefont {Tao}\ \bibnamefont
  {Liu}}, \bibinfo {author} {\bibfnamefont {Yu-Ran}\ \bibnamefont {Zhang}},
  \bibinfo {author} {\bibfnamefont {Qing}\ \bibnamefont {Ai}}, \bibinfo
  {author} {\bibfnamefont {Zongping}\ \bibnamefont {Gong}}, \bibinfo {author}
  {\bibfnamefont {Kohei}\ \bibnamefont {Kawabata}}, \bibinfo {author}
  {\bibfnamefont {Masahito}\ \bibnamefont {Ueda}}, \ and\ \bibinfo {author}
  {\bibfnamefont {Franco}\ \bibnamefont {Nori}},\ }\bibfield  {title} {\enquote
  {\bibinfo {title} {Second-order topological phases in non-hermitian
  systems},}\ }\href {\doibase 10.1103/PhysRevLett.122.076801} {\bibfield
  {journal} {\bibinfo  {journal} {Phys. Rev. Lett.}\ }\textbf {\bibinfo
  {volume} {122}},\ \bibinfo {pages} {076801} (\bibinfo {year}
  {2019})}\BibitemShut {NoStop}%
\bibitem [{\citenamefont {Lieu}(2018{\natexlab{a}})}]{article36}%
  \BibitemOpen
  \bibfield  {author} {\bibinfo {author} {\bibfnamefont {Simon}\ \bibnamefont
  {Lieu}},\ }\bibfield  {title} {\enquote {\bibinfo {title} {Topological phases
  in the non-hermitian su-schrieffer-heeger model},}\ }\href {\doibase
  10.1103/PhysRevB.97.045106} {\bibfield  {journal} {\bibinfo  {journal} {Phys.
  Rev. B}\ }\textbf {\bibinfo {volume} {97}},\ \bibinfo {pages} {045106}
  (\bibinfo {year} {2018}{\natexlab{a}})}\BibitemShut {NoStop}%
\bibitem [{\citenamefont {Gong}\ \emph {et~al.}(2018)\citenamefont {Gong},
  \citenamefont {Ashida}, \citenamefont {Kawabata}, \citenamefont {Takasan},
  \citenamefont {Higashikawa},\ and\ \citenamefont {Ueda}}]{article37}%
  \BibitemOpen
  \bibfield  {author} {\bibinfo {author} {\bibfnamefont {Zongping}\
  \bibnamefont {Gong}}, \bibinfo {author} {\bibfnamefont {Yuto}\ \bibnamefont
  {Ashida}}, \bibinfo {author} {\bibfnamefont {Kohei}\ \bibnamefont
  {Kawabata}}, \bibinfo {author} {\bibfnamefont {Kazuaki}\ \bibnamefont
  {Takasan}}, \bibinfo {author} {\bibfnamefont {Sho}\ \bibnamefont
  {Higashikawa}}, \ and\ \bibinfo {author} {\bibfnamefont {Masahito}\
  \bibnamefont {Ueda}},\ }\bibfield  {title} {\enquote {\bibinfo {title}
  {Topological phases of non-hermitian systems},}\ }\href {\doibase
  10.1103/PhysRevX.8.031079} {\bibfield  {journal} {\bibinfo  {journal} {Phys.
  Rev. X}\ }\textbf {\bibinfo {volume} {8}},\ \bibinfo {pages} {031079}
  (\bibinfo {year} {2018})}\BibitemShut {NoStop}%
\bibitem [{\citenamefont {Ghatak}\ and\ \citenamefont {Das}(2019)}]{article38}%
  \BibitemOpen
  \bibfield  {author} {\bibinfo {author} {\bibfnamefont {Ananya}\ \bibnamefont
  {Ghatak}}\ and\ \bibinfo {author} {\bibfnamefont {Tanmoy}\ \bibnamefont
  {Das}},\ }\bibfield  {title} {\enquote {\bibinfo {title} {New topological
  invariants in non-hermitian systems},}\ }\href {\doibase
  10.1088/1361-648x/ab11b3} {\bibfield  {journal} {\bibinfo  {journal} {J.
  Phys. Condens. Matter}\ }\textbf {\bibinfo {volume} {31}},\ \bibinfo {pages}
  {263001} (\bibinfo {year} {2019})}\BibitemShut {NoStop}%
\bibitem [{\citenamefont {Yin}\ \emph {et~al.}(2018)\citenamefont {Yin},
  \citenamefont {Jiang}, \citenamefont {Li}, \citenamefont {L\"u},\ and\
  \citenamefont {Chen}}]{article39}%
  \BibitemOpen
  \bibfield  {author} {\bibinfo {author} {\bibfnamefont {Chuanhao}\
  \bibnamefont {Yin}}, \bibinfo {author} {\bibfnamefont {Hui}\ \bibnamefont
  {Jiang}}, \bibinfo {author} {\bibfnamefont {Linhu}\ \bibnamefont {Li}},
  \bibinfo {author} {\bibfnamefont {Rong}\ \bibnamefont {L\"u}}, \ and\
  \bibinfo {author} {\bibfnamefont {Shu}\ \bibnamefont {Chen}},\ }\bibfield
  {title} {\enquote {\bibinfo {title} {Geometrical meaning of winding number
  and its characterization of topological phases in one-dimensional chiral
  non-hermitian systems},}\ }\href {\doibase 10.1103/PhysRevA.97.052115}
  {\bibfield  {journal} {\bibinfo  {journal} {Phys. Rev. A}\ }\textbf {\bibinfo
  {volume} {97}},\ \bibinfo {pages} {052115} (\bibinfo {year}
  {2018})}\BibitemShut {NoStop}%
\bibitem [{\citenamefont {Esaki}\ \emph {et~al.}(2011)\citenamefont {Esaki},
  \citenamefont {Sato}, \citenamefont {Hasebe},\ and\ \citenamefont
  {Kohmoto}}]{article40}%
  \BibitemOpen
  \bibfield  {author} {\bibinfo {author} {\bibfnamefont {Kenta}\ \bibnamefont
  {Esaki}}, \bibinfo {author} {\bibfnamefont {Masatoshi}\ \bibnamefont {Sato}},
  \bibinfo {author} {\bibfnamefont {Kazuki}\ \bibnamefont {Hasebe}}, \ and\
  \bibinfo {author} {\bibfnamefont {Mahito}\ \bibnamefont {Kohmoto}},\
  }\bibfield  {title} {\enquote {\bibinfo {title} {Edge states and topological
  phases in non-hermitian systems},}\ }\href {\doibase
  10.1103/PhysRevB.84.205128} {\bibfield  {journal} {\bibinfo  {journal} {Phys.
  Rev. B}\ }\textbf {\bibinfo {volume} {84}},\ \bibinfo {pages} {205128}
  (\bibinfo {year} {2011})}\BibitemShut {NoStop}%
\bibitem [{\citenamefont {Song}\ \emph
  {et~al.}(2019{\natexlab{a}})\citenamefont {Song}, \citenamefont {Yao},\ and\
  \citenamefont {Wang}}]{article41}%
  \BibitemOpen
  \bibfield  {author} {\bibinfo {author} {\bibfnamefont {Fei}\ \bibnamefont
  {Song}}, \bibinfo {author} {\bibfnamefont {Shunyu}\ \bibnamefont {Yao}}, \
  and\ \bibinfo {author} {\bibfnamefont {Zhong}\ \bibnamefont {Wang}},\
  }\bibfield  {title} {\enquote {\bibinfo {title} {Non-hermitian topological
  invariants in real space},}\ }\href {\doibase 10.1103/PhysRevLett.123.246801}
  {\bibfield  {journal} {\bibinfo  {journal} {Phys. Rev. Lett.}\ }\textbf
  {\bibinfo {volume} {123}},\ \bibinfo {pages} {246801} (\bibinfo {year}
  {2019}{\natexlab{a}})}\BibitemShut {NoStop}%
\bibitem [{\citenamefont {Yang}\ \emph {et~al.}()\citenamefont {Yang},
  \citenamefont {Cao},\ and\ \citenamefont {Zhai}}]{article42}%
  \BibitemOpen
  \bibfield  {author} {\bibinfo {author} {\bibfnamefont {Xiaoseng}\
  \bibnamefont {Yang}}, \bibinfo {author} {\bibfnamefont {Yang}\ \bibnamefont
  {Cao}}, \ and\ \bibinfo {author} {\bibfnamefont {Yunjia}\ \bibnamefont
  {Zhai}},\ }\bibfield  {title} {\enquote {\bibinfo {title} {Non-hermitian weyl
  semimetals: Non-hermitian skin eﬀect and non-bloch bulk-boundary
  correspondence},}\ }\href {https://arxiv.org/abs/1904.02492} {\bibinfo
  {journal} {arXiv: 1911.02492}\ }\BibitemShut {NoStop}%
\bibitem [{\citenamefont {Ezawa}(2019)}]{article48}%
  \BibitemOpen
\bibfield  {journal} {  }\bibfield  {author} {\bibinfo {author} {\bibfnamefont
  {Motohiko}\ \bibnamefont {Ezawa}},\ }\bibfield  {title} {\enquote {\bibinfo
  {title} {Braiding of majorana-like corner states in electric circuits and its
  non-hermitian generalization},}\ }\href {\doibase
  10.1103/PhysRevB.100.045407} {\bibfield  {journal} {\bibinfo  {journal}
  {Phys. Rev. B}\ }\textbf {\bibinfo {volume} {100}},\ \bibinfo {pages}
  {045407} (\bibinfo {year} {2019})}\BibitemShut {NoStop}%
\bibitem [{\citenamefont {Ge}\ \emph {et~al.}(2019)\citenamefont {Ge},
  \citenamefont {Zhang}, \citenamefont {Liu}, \citenamefont {Li}, \citenamefont
  {Fan},\ and\ \citenamefont {Nori}}]{article49}%
  \BibitemOpen
  \bibfield  {author} {\bibinfo {author} {\bibfnamefont {Zi-Yong}\ \bibnamefont
  {Ge}}, \bibinfo {author} {\bibfnamefont {Yu-Ran}\ \bibnamefont {Zhang}},
  \bibinfo {author} {\bibfnamefont {Tao}\ \bibnamefont {Liu}}, \bibinfo
  {author} {\bibfnamefont {Si-Wen}\ \bibnamefont {Li}}, \bibinfo {author}
  {\bibfnamefont {Heng}\ \bibnamefont {Fan}}, \ and\ \bibinfo {author}
  {\bibfnamefont {Franco}\ \bibnamefont {Nori}},\ }\bibfield  {title} {\enquote
  {\bibinfo {title} {Topological band theory for non-hermitian systems from the
  dirac equation},}\ }\href {\doibase 10.1103/PhysRevB.100.054105} {\bibfield
  {journal} {\bibinfo  {journal} {Phys. Rev. B}\ }\textbf {\bibinfo {volume}
  {100}},\ \bibinfo {pages} {054105} (\bibinfo {year} {2019})}\BibitemShut
  {NoStop}%
\bibitem [{\citenamefont {Jiang}\ \emph {et~al.}(2019)\citenamefont {Jiang},
  \citenamefont {Lang}, \citenamefont {Yang}, \citenamefont {Zhu},\ and\
  \citenamefont {Chen}}]{article50}%
  \BibitemOpen
  \bibfield  {author} {\bibinfo {author} {\bibfnamefont {Hui}\ \bibnamefont
  {Jiang}}, \bibinfo {author} {\bibfnamefont {Li-Jun}\ \bibnamefont {Lang}},
  \bibinfo {author} {\bibfnamefont {Chao}\ \bibnamefont {Yang}}, \bibinfo
  {author} {\bibfnamefont {Shi-Liang}\ \bibnamefont {Zhu}}, \ and\ \bibinfo
  {author} {\bibfnamefont {Shu}\ \bibnamefont {Chen}},\ }\bibfield  {title}
  {\enquote {\bibinfo {title} {Interplay of non-hermitian skin effects and
  anderson localization in nonreciprocal quasiperiodic lattices},}\ }\href
  {\doibase 10.1103/PhysRevB.100.054301} {\bibfield  {journal} {\bibinfo
  {journal} {Phys. Rev. B}\ }\textbf {\bibinfo {volume} {100}},\ \bibinfo
  {pages} {054301} (\bibinfo {year} {2019})}\BibitemShut {NoStop}%
\bibitem [{\citenamefont {Lee}\ \emph {et~al.}(2019)\citenamefont {Lee},
  \citenamefont {Li},\ and\ \citenamefont {Gong}}]{article51}%
  \BibitemOpen
  \bibfield  {author} {\bibinfo {author} {\bibfnamefont {Ching~Hua}\
  \bibnamefont {Lee}}, \bibinfo {author} {\bibfnamefont {Linhu}\ \bibnamefont
  {Li}}, \ and\ \bibinfo {author} {\bibfnamefont {Jiangbin}\ \bibnamefont
  {Gong}},\ }\bibfield  {title} {\enquote {\bibinfo {title} {Hybrid
  higher-order skin-topological modes in nonreciprocal systems},}\ }\href
  {\doibase 10.1103/PhysRevLett.123.016805} {\bibfield  {journal} {\bibinfo
  {journal} {Phys. Rev. Lett.}\ }\textbf {\bibinfo {volume} {123}},\ \bibinfo
  {pages} {016805} (\bibinfo {year} {2019})}\BibitemShut {NoStop}%
\bibitem [{\citenamefont {Okuma}\ and\ \citenamefont {Sato}(2019)}]{article52}%
  \BibitemOpen
  \bibfield  {author} {\bibinfo {author} {\bibfnamefont {Nobuyuki}\
  \bibnamefont {Okuma}}\ and\ \bibinfo {author} {\bibfnamefont {Masatoshi}\
  \bibnamefont {Sato}},\ }\bibfield  {title} {\enquote {\bibinfo {title}
  {Topological phase transition driven by infinitesimal instability: Majorana
  fermions in non-hermitian spintronics},}\ }\href {\doibase
  10.1103/PhysRevLett.123.097701} {\bibfield  {journal} {\bibinfo  {journal}
  {Phys. Rev. Lett.}\ }\textbf {\bibinfo {volume} {123}},\ \bibinfo {pages}
  {097701} (\bibinfo {year} {2019})}\BibitemShut {NoStop}%
\bibitem [{\citenamefont {Song}\ \emph
  {et~al.}(2019{\natexlab{b}})\citenamefont {Song}, \citenamefont {Yao},\ and\
  \citenamefont {Wang}}]{article53}%
  \BibitemOpen
  \bibfield  {author} {\bibinfo {author} {\bibfnamefont {Fei}\ \bibnamefont
  {Song}}, \bibinfo {author} {\bibfnamefont {Shunyu}\ \bibnamefont {Yao}}, \
  and\ \bibinfo {author} {\bibfnamefont {Zhong}\ \bibnamefont {Wang}},\
  }\bibfield  {title} {\enquote {\bibinfo {title} {Non-hermitian skin effect
  and chiral damping in open quantum systems},}\ }\href {\doibase
  10.1103/PhysRevLett.123.170401} {\bibfield  {journal} {\bibinfo  {journal}
  {Phys. Rev. Lett.}\ }\textbf {\bibinfo {volume} {123}},\ \bibinfo {pages}
  {170401} (\bibinfo {year} {2019}{\natexlab{b}})}\BibitemShut {NoStop}%
\bibitem [{\citenamefont {Zhou}\ and\ \citenamefont {Lee}(2019)}]{article29}%
  \BibitemOpen
  \bibfield  {author} {\bibinfo {author} {\bibfnamefont {Hengyun}\ \bibnamefont
  {Zhou}}\ and\ \bibinfo {author} {\bibfnamefont {Jong~Yeon}\ \bibnamefont
  {Lee}},\ }\bibfield  {title} {\enquote {\bibinfo {title} {Periodic table for
  topological bands with non-hermitian symmetries},}\ }\href {\doibase
  10.1103/PhysRevB.99.235112} {\bibfield  {journal} {\bibinfo  {journal} {Phys.
  Rev. B}\ }\textbf {\bibinfo {volume} {99}},\ \bibinfo {pages} {235112}
  (\bibinfo {year} {2019})}\BibitemShut {NoStop}%
\bibitem [{\citenamefont {Kawabata}\ \emph
  {et~al.}(2019{\natexlab{a}})\citenamefont {Kawabata}, \citenamefont
  {Shiozaki}, \citenamefont {Ueda},\ and\ \citenamefont {Sato}}]{article30}%
  \BibitemOpen
  \bibfield  {author} {\bibinfo {author} {\bibfnamefont {Kohei}\ \bibnamefont
  {Kawabata}}, \bibinfo {author} {\bibfnamefont {Ken}\ \bibnamefont
  {Shiozaki}}, \bibinfo {author} {\bibfnamefont {Masahito}\ \bibnamefont
  {Ueda}}, \ and\ \bibinfo {author} {\bibfnamefont {Masatoshi}\ \bibnamefont
  {Sato}},\ }\bibfield  {title} {\enquote {\bibinfo {title} {Symmetry and
  topology in non-hermitian physics},}\ }\href {\doibase
  10.1103/PhysRevX.9.041015} {\bibfield  {journal} {\bibinfo  {journal} {Phys.
  Rev. X}\ }\textbf {\bibinfo {volume} {9}},\ \bibinfo {pages} {041015}
  (\bibinfo {year} {2019}{\natexlab{a}})}\BibitemShut {NoStop}%
\bibitem [{\citenamefont {Kawabata}\ \emph {et~al.}(2018)\citenamefont
  {Kawabata}, \citenamefont {Higashikawa}, \citenamefont {Gong}, \citenamefont
  {Ashida},\ and\ \citenamefont {Ueda}}]{article31}%
  \BibitemOpen
  \bibfield  {author} {\bibinfo {author} {\bibfnamefont {Kohei}\ \bibnamefont
  {Kawabata}}, \bibinfo {author} {\bibfnamefont {Sho}\ \bibnamefont
  {Higashikawa}}, \bibinfo {author} {\bibfnamefont {Zongping}\ \bibnamefont
  {Gong}}, \bibinfo {author} {\bibfnamefont {Yuto}\ \bibnamefont {Ashida}}, \
  and\ \bibinfo {author} {\bibfnamefont {Masahito}\ \bibnamefont {Ueda}},\
  }\bibfield  {title} {\enquote {\bibinfo {title} {Topological unification of
  time-reversal and particle-hole symmetries in non-hermitian physics},}\
  }\href {https://doi.org/10.1038/s41467-018-08254-y} {\bibfield  {journal}
  {\bibinfo  {journal} {Nat. Commun.}\ }\textbf {\bibinfo {volume} {10}},\
  \bibinfo {pages} {297} (\bibinfo {year} {2018})}\BibitemShut {NoStop}%
\bibitem [{\citenamefont {Xi}\ \emph {et~al.}()\citenamefont {Xi},
  \citenamefont {Zhang}, \citenamefont {Gu},\ and\ \citenamefont
  {Chen}}]{article32}%
  \BibitemOpen
  \bibfield  {author} {\bibinfo {author} {\bibfnamefont {Wenjie}\ \bibnamefont
  {Xi}}, \bibinfo {author} {\bibfnamefont {Zhi-Hao}\ \bibnamefont {Zhang}},
  \bibinfo {author} {\bibfnamefont {Zheng-Cheng}\ \bibnamefont {Gu}}, \ and\
  \bibinfo {author} {\bibfnamefont {Wei-Qiang}\ \bibnamefont {Chen}},\
  }\bibfield  {title} {\enquote {\bibinfo {title} {Classification of
  topological phases in one dimensional interacting non-hermitian systems and
  emergent unitarity},}\ }\href {https://arxiv.org/abs/1911.01590} {\bibinfo
  {journal} {arXiv: 1911.01590}\ }\BibitemShut {NoStop}%
\bibitem [{\citenamefont {Hatano}\ and\ \citenamefont
  {Nelson}(1996)}]{article54}%
  \BibitemOpen
\bibfield  {journal} {  }\bibfield  {author} {\bibinfo {author} {\bibfnamefont
  {Naomichi}\ \bibnamefont {Hatano}}\ and\ \bibinfo {author} {\bibfnamefont
  {David~R.}\ \bibnamefont {Nelson}},\ }\bibfield  {title} {\enquote {\bibinfo
  {title} {Localization transitions in non-hermitian quantum mechanics},}\
  }\href {\doibase 10.1103/PhysRevLett.77.570} {\bibfield  {journal} {\bibinfo
  {journal} {Phys. Rev. Lett.}\ }\textbf {\bibinfo {volume} {77}},\ \bibinfo
  {pages} {570--573} (\bibinfo {year} {1996})}\BibitemShut {NoStop}%
\bibitem [{\citenamefont {McDonald}\ \emph {et~al.}(2018)\citenamefont
  {McDonald}, \citenamefont {Pereg-Barnea},\ and\ \citenamefont
  {Clerk}}]{article57}%
  \BibitemOpen
  \bibfield  {author} {\bibinfo {author} {\bibfnamefont {A.}~\bibnamefont
  {McDonald}}, \bibinfo {author} {\bibfnamefont {T.}~\bibnamefont
  {Pereg-Barnea}}, \ and\ \bibinfo {author} {\bibfnamefont {A.~A.}\
  \bibnamefont {Clerk}},\ }\bibfield  {title} {\enquote {\bibinfo {title}
  {Phase-dependent chiral transport and effective non-hermitian dynamics in a
  bosonic kitaev-majorana chain},}\ }\href {\doibase 10.1103/PhysRevX.8.041031}
  {\bibfield  {journal} {\bibinfo  {journal} {Phys. Rev. X}\ }\textbf {\bibinfo
  {volume} {8}},\ \bibinfo {pages} {041031} (\bibinfo {year}
  {2018})}\BibitemShut {NoStop}%
\bibitem [{\citenamefont {Silveirinha}(2019{\natexlab{a}})}]{article59}%
  \BibitemOpen
  \bibfield  {author} {\bibinfo {author} {\bibfnamefont {M\'ario~G.}\
  \bibnamefont {Silveirinha}},\ }\bibfield  {title} {\enquote {\bibinfo {title}
  {Topological theory of non-hermitian photonic systems},}\ }\href {\doibase
  10.1103/PhysRevB.99.125155} {\bibfield  {journal} {\bibinfo  {journal} {Phys.
  Rev. B}\ }\textbf {\bibinfo {volume} {99}},\ \bibinfo {pages} {125155}
  (\bibinfo {year} {2019}{\natexlab{a}})}\BibitemShut {NoStop}%
\bibitem [{\citenamefont {Kawabata}\ \emph
  {et~al.}(2019{\natexlab{b}})\citenamefont {Kawabata}, \citenamefont
  {Bessho},\ and\ \citenamefont {Sato}}]{article60}%
  \BibitemOpen
  \bibfield  {author} {\bibinfo {author} {\bibfnamefont {Kohei}\ \bibnamefont
  {Kawabata}}, \bibinfo {author} {\bibfnamefont {Takumi}\ \bibnamefont
  {Bessho}}, \ and\ \bibinfo {author} {\bibfnamefont {Masatoshi}\ \bibnamefont
  {Sato}},\ }\bibfield  {title} {\enquote {\bibinfo {title} {Classification of
  exceptional points and non-hermitian topological semimetals},}\ }\href
  {\doibase 10.1103/PhysRevLett.123.066405} {\bibfield  {journal} {\bibinfo
  {journal} {Phys. Rev. Lett.}\ }\textbf {\bibinfo {volume} {123}},\ \bibinfo
  {pages} {066405} (\bibinfo {year} {2019}{\natexlab{b}})}\BibitemShut
  {NoStop}%
\bibitem [{\citenamefont {Rui}\ \emph {et~al.}(2019)\citenamefont {Rui},
  \citenamefont {Zhao},\ and\ \citenamefont {Schnyder}}]{article62}%
  \BibitemOpen
  \bibfield  {author} {\bibinfo {author} {\bibfnamefont {W.~B.}\ \bibnamefont
  {Rui}}, \bibinfo {author} {\bibfnamefont {Y.~X.}\ \bibnamefont {Zhao}}, \
  and\ \bibinfo {author} {\bibfnamefont {Andreas~P.}\ \bibnamefont
  {Schnyder}},\ }\bibfield  {title} {\enquote {\bibinfo {title} {Topology and
  exceptional points of massive dirac models with generic non-hermitian
  perturbations},}\ }\href {\doibase 10.1103/PhysRevB.99.241110} {\bibfield
  {journal} {\bibinfo  {journal} {Phys. Rev. B}\ }\textbf {\bibinfo {volume}
  {99}},\ \bibinfo {pages} {241110} (\bibinfo {year} {2019})}\BibitemShut
  {NoStop}%
\bibitem [{\citenamefont {Ozcakmakli~Turker}\ and\ \citenamefont
  {Yuce}(2019)}]{article65}%
  \BibitemOpen
  \bibfield  {author} {\bibinfo {author} {\bibfnamefont {Z.}~\bibnamefont
  {Ozcakmakli~Turker}}\ and\ \bibinfo {author} {\bibfnamefont {C.}~\bibnamefont
  {Yuce}},\ }\bibfield  {title} {\enquote {\bibinfo {title} {Open and closed
  boundaries in non-hermitian topological systems},}\ }\href {\doibase
  10.1103/PhysRevA.99.022127} {\bibfield  {journal} {\bibinfo  {journal} {Phys.
  Rev. A}\ }\textbf {\bibinfo {volume} {99}},\ \bibinfo {pages} {022127}
  (\bibinfo {year} {2019})}\BibitemShut {NoStop}%
\bibitem [{\citenamefont {Harari}\ \emph {et~al.}(2018)\citenamefont {Harari},
  \citenamefont {Bandres}, \citenamefont {Lumer}, \citenamefont {Rechtsman},
  \citenamefont {Chong}, \citenamefont {Khajavikhan}, \citenamefont
  {Christodoulides},\ and\ \citenamefont {Segev}}]{article67}%
  \BibitemOpen
  \bibfield  {author} {\bibinfo {author} {\bibfnamefont {Gal}\ \bibnamefont
  {Harari}}, \bibinfo {author} {\bibfnamefont {Miguel~A.}\ \bibnamefont
  {Bandres}}, \bibinfo {author} {\bibfnamefont {Yaakov}\ \bibnamefont {Lumer}},
  \bibinfo {author} {\bibfnamefont {Mikael~C.}\ \bibnamefont {Rechtsman}},
  \bibinfo {author} {\bibfnamefont {Y.~D.}\ \bibnamefont {Chong}}, \bibinfo
  {author} {\bibfnamefont {Mercedeh}\ \bibnamefont {Khajavikhan}}, \bibinfo
  {author} {\bibfnamefont {Demetrios~N.}\ \bibnamefont {Christodoulides}}, \
  and\ \bibinfo {author} {\bibfnamefont {Mordechai}\ \bibnamefont {Segev}},\
  }\bibfield  {title} {\enquote {\bibinfo {title} {Topological insulator laser:
  Theory},}\ }\href {https://science.sciencemag.org/content/359/6381/eaar4003}
  {\bibfield  {journal} {\bibinfo  {journal} {Science}\ }\textbf {\bibinfo
  {volume} {359}},\ \bibinfo {pages} {eaar4003} (\bibinfo {year}
  {2018})}\BibitemShut {NoStop}%
\bibitem [{\citenamefont {Zhu}\ \emph {et~al.}(2014)\citenamefont {Zhu},
  \citenamefont {L\"u},\ and\ \citenamefont {Chen}}]{article68}%
  \BibitemOpen
  \bibfield  {author} {\bibinfo {author} {\bibfnamefont {Baogang}\ \bibnamefont
  {Zhu}}, \bibinfo {author} {\bibfnamefont {Rong}\ \bibnamefont {L\"u}}, \ and\
  \bibinfo {author} {\bibfnamefont {Shu}\ \bibnamefont {Chen}},\ }\bibfield
  {title} {\enquote {\bibinfo {title} {$\mathcal{PT}$ symmetry in the
  non-hermitian su-schrieffer-heeger model with complex boundary potentials},}\
  }\href {\doibase 10.1103/PhysRevA.89.062102} {\bibfield  {journal} {\bibinfo
  {journal} {Phys. Rev. A}\ }\textbf {\bibinfo {volume} {89}},\ \bibinfo
  {pages} {062102} (\bibinfo {year} {2014})}\BibitemShut {NoStop}%
\bibitem [{\citenamefont {Lieu}(2018{\natexlab{b}})}]{article69}%
  \BibitemOpen
  \bibfield  {author} {\bibinfo {author} {\bibfnamefont {Simon}\ \bibnamefont
  {Lieu}},\ }\bibfield  {title} {\enquote {\bibinfo {title} {Topological
  symmetry classes for non-hermitian models and connections to the bosonic
  bogoliubov--de gennes equation},}\ }\href {\doibase
  10.1103/PhysRevB.98.115135} {\bibfield  {journal} {\bibinfo  {journal} {Phys.
  Rev. B}\ }\textbf {\bibinfo {volume} {98}},\ \bibinfo {pages} {115135}
  (\bibinfo {year} {2018}{\natexlab{b}})}\BibitemShut {NoStop}%
\bibitem [{\citenamefont {Yuce}(2016)}]{article70}%
  \BibitemOpen
  \bibfield  {author} {\bibinfo {author} {\bibfnamefont {C.}~\bibnamefont
  {Yuce}},\ }\bibfield  {title} {\enquote {\bibinfo {title} {Majorana edge
  modes with gain and loss},}\ }\href {\doibase 10.1103/PhysRevA.93.062130}
  {\bibfield  {journal} {\bibinfo  {journal} {Phys. Rev. A}\ }\textbf {\bibinfo
  {volume} {93}},\ \bibinfo {pages} {062130} (\bibinfo {year}
  {2016})}\BibitemShut {NoStop}%
\bibitem [{\citenamefont {Klett}\ \emph {et~al.}(2017)\citenamefont {Klett},
  \citenamefont {Cartarius}, \citenamefont {Dast}, \citenamefont {Main},\ and\
  \citenamefont {Wunner}}]{article75}%
  \BibitemOpen
  \bibfield  {author} {\bibinfo {author} {\bibfnamefont {Marcel}\ \bibnamefont
  {Klett}}, \bibinfo {author} {\bibfnamefont {Holger}\ \bibnamefont
  {Cartarius}}, \bibinfo {author} {\bibfnamefont {Dennis}\ \bibnamefont
  {Dast}}, \bibinfo {author} {\bibfnamefont {J\"org}\ \bibnamefont {Main}}, \
  and\ \bibinfo {author} {\bibfnamefont {G\"unter}\ \bibnamefont {Wunner}},\
  }\bibfield  {title} {\enquote {\bibinfo {title} {Relation between
  $\mathcal{PT}$-symmetry breaking and topologically nontrivial phases in the
  su-schrieffer-heeger and kitaev models},}\ }\href {\doibase
  10.1103/PhysRevA.95.053626} {\bibfield  {journal} {\bibinfo  {journal} {Phys.
  Rev. A}\ }\textbf {\bibinfo {volume} {95}},\ \bibinfo {pages} {053626}
  (\bibinfo {year} {2017})}\BibitemShut {NoStop}%
\bibitem [{\citenamefont {Zeuner}\ \emph {et~al.}(2015)\citenamefont {Zeuner},
  \citenamefont {Rechtsman}, \citenamefont {Plotnik}, \citenamefont {Lumer},
  \citenamefont {Nolte}, \citenamefont {Rudner}, \citenamefont {Segev},\ and\
  \citenamefont {Szameit}}]{article77}%
  \BibitemOpen
  \bibfield  {author} {\bibinfo {author} {\bibfnamefont {Julia~M.}\
  \bibnamefont {Zeuner}}, \bibinfo {author} {\bibfnamefont {Mikael~C.}\
  \bibnamefont {Rechtsman}}, \bibinfo {author} {\bibfnamefont {Yonatan}\
  \bibnamefont {Plotnik}}, \bibinfo {author} {\bibfnamefont {Yaakov}\
  \bibnamefont {Lumer}}, \bibinfo {author} {\bibfnamefont {Stefan}\
  \bibnamefont {Nolte}}, \bibinfo {author} {\bibfnamefont {Mark~S.}\
  \bibnamefont {Rudner}}, \bibinfo {author} {\bibfnamefont {Mordechai}\
  \bibnamefont {Segev}}, \ and\ \bibinfo {author} {\bibfnamefont {Alexander}\
  \bibnamefont {Szameit}},\ }\bibfield  {title} {\enquote {\bibinfo {title}
  {Observation of a topological transition in the bulk of a non-hermitian
  system},}\ }\href {\doibase 10.1103/PhysRevLett.115.040402} {\bibfield
  {journal} {\bibinfo  {journal} {Phys. Rev. Lett.}\ }\textbf {\bibinfo
  {volume} {115}},\ \bibinfo {pages} {040402} (\bibinfo {year}
  {2015})}\BibitemShut {NoStop}%
\bibitem [{\citenamefont {Zeng}\ \emph {et~al.}(2020)\citenamefont {Zeng},
  \citenamefont {Yang},\ and\ \citenamefont {Xu}}]{article85}%
  \BibitemOpen
  \bibfield  {author} {\bibinfo {author} {\bibfnamefont {Qi-Bo}\ \bibnamefont
  {Zeng}}, \bibinfo {author} {\bibfnamefont {Yan-Bin}\ \bibnamefont {Yang}}, \
  and\ \bibinfo {author} {\bibfnamefont {Yong}\ \bibnamefont {Xu}},\ }\bibfield
   {title} {\enquote {\bibinfo {title} {Topological phases in non-hermitian
  aubry-andr\'e-harper models},}\ }\href {\doibase 10.1103/PhysRevB.101.020201}
  {\bibfield  {journal} {\bibinfo  {journal} {Phys. Rev. B}\ }\textbf {\bibinfo
  {volume} {101}},\ \bibinfo {pages} {020201} (\bibinfo {year}
  {2020})}\BibitemShut {NoStop}%
\bibitem [{\citenamefont {San-Jose}\ \emph {et~al.}(2016)\citenamefont
  {San-Jose}, \citenamefont {Cayao}, \citenamefont {Prada},\ and\ \citenamefont
  {Aguado}}]{article87}%
  \BibitemOpen
  \bibfield  {author} {\bibinfo {author} {\bibfnamefont {Pablo}\ \bibnamefont
  {San-Jose}}, \bibinfo {author} {\bibfnamefont {Jorge}\ \bibnamefont {Cayao}},
  \bibinfo {author} {\bibfnamefont {Elsa}\ \bibnamefont {Prada}}, \ and\
  \bibinfo {author} {\bibfnamefont {Ramon}\ \bibnamefont {Aguado}},\ }\bibfield
   {title} {\enquote {\bibinfo {title} {Majorana bound states from exceptional
  points in non-topological superconductors},}\ }\href {\doibase
  10.1038/srep21427} {\bibfield  {journal} {\bibinfo  {journal} {Sci. Rep.}\
  }\textbf {\bibinfo {volume} {6}},\ \bibinfo {pages} {21427} (\bibinfo {year}
  {2016})}\BibitemShut {NoStop}%
\bibitem [{\citenamefont {Avila}\ \emph {et~al.}(2019)\citenamefont {Avila},
  \citenamefont {Peñaranda}, \citenamefont {Prada}, \citenamefont {San-Jose},\
  and\ \citenamefont {Aguado}}]{article88}%
  \BibitemOpen
  \bibfield  {author} {\bibinfo {author} {\bibfnamefont {J.}~\bibnamefont
  {Avila}}, \bibinfo {author} {\bibfnamefont {Fernando}\ \bibnamefont
  {Peñaranda}}, \bibinfo {author} {\bibfnamefont {Elsa}\ \bibnamefont
  {Prada}}, \bibinfo {author} {\bibfnamefont {Pablo}\ \bibnamefont {San-Jose}},
  \ and\ \bibinfo {author} {\bibfnamefont {Ramon}\ \bibnamefont {Aguado}},\
  }\bibfield  {title} {\enquote {\bibinfo {title} {Non-hermitian topology as a
  unifying framework for the andreev versus majorana states controversy},}\
  }\href {https://doi.org/10.1038/s42005-019-0231-8} {\bibfield  {journal}
  {\bibinfo  {journal} {Commun. Phys.}\ }\textbf {\bibinfo {volume} {2}},\
  \bibinfo {pages} {133} (\bibinfo {year} {2019})}\BibitemShut {NoStop}%
\bibitem [{\citenamefont {Silveirinha}(2019{\natexlab{b}})}]{article89}%
  \BibitemOpen
  \bibfield  {author} {\bibinfo {author} {\bibfnamefont {M\'ario~G.}\
  \bibnamefont {Silveirinha}},\ }\bibfield  {title} {\enquote {\bibinfo {title}
  {Topological theory of non-hermitian photonic systems},}\ }\href {\doibase
  10.1103/PhysRevB.99.125155} {\bibfield  {journal} {\bibinfo  {journal} {Phys.
  Rev. B}\ }\textbf {\bibinfo {volume} {99}},\ \bibinfo {pages} {125155}
  (\bibinfo {year} {2019}{\natexlab{b}})}\BibitemShut {NoStop}%
\bibitem [{\citenamefont {Yamamoto}\ \emph {et~al.}(2019)\citenamefont
  {Yamamoto}, \citenamefont {Nakagawa}, \citenamefont {Adachi}, \citenamefont
  {Takasan}, \citenamefont {Ueda},\ and\ \citenamefont {Kawakami}}]{article90}%
  \BibitemOpen
  \bibfield  {author} {\bibinfo {author} {\bibfnamefont {Kazuki}\ \bibnamefont
  {Yamamoto}}, \bibinfo {author} {\bibfnamefont {Masaya}\ \bibnamefont
  {Nakagawa}}, \bibinfo {author} {\bibfnamefont {Kyosuke}\ \bibnamefont
  {Adachi}}, \bibinfo {author} {\bibfnamefont {Kazuaki}\ \bibnamefont
  {Takasan}}, \bibinfo {author} {\bibfnamefont {Masahito}\ \bibnamefont
  {Ueda}}, \ and\ \bibinfo {author} {\bibfnamefont {Norio}\ \bibnamefont
  {Kawakami}},\ }\bibfield  {title} {\enquote {\bibinfo {title} {Theory of
  non-hermitian fermionic superfluidity with a complex-valued interaction},}\
  }\href {\doibase 10.1103/PhysRevLett.123.123601} {\bibfield  {journal}
  {\bibinfo  {journal} {Phys. Rev. Lett.}\ }\textbf {\bibinfo {volume} {123}},\
  \bibinfo {pages} {123601} (\bibinfo {year} {2019})}\BibitemShut {NoStop}%
\bibitem [{sup()}]{sup}%
  \BibitemOpen
  \href@noop {} {\enquote {\bibinfo {title} {See supplemental material for
  details of calculation.}}\ }\BibitemShut {NoStop}%
\bibitem [{\citenamefont {Kadanoff}(1977)}]{article94}%
  \BibitemOpen
  \bibfield  {author} {\bibinfo {author} {\bibfnamefont {Leo~P.}\ \bibnamefont
  {Kadanoff}},\ }\bibfield  {title} {\enquote {\bibinfo {title} {The
  application of renormalization group techniques to quarks and strings},}\
  }\href {\doibase 10.1103/RevModPhys.49.267} {\bibfield  {journal} {\bibinfo
  {journal} {Rev. Mod. Phys.}\ }\textbf {\bibinfo {volume} {49}},\ \bibinfo
  {pages} {267--296} (\bibinfo {year} {1977})}\BibitemShut {NoStop}%
\bibitem [{\citenamefont {Kadanoff}\ \emph {et~al.}(1967)\citenamefont
  {Kadanoff}, \citenamefont {G\"otze}, \citenamefont {Hamblen}, \citenamefont
  {Hecht}, \citenamefont {Lewis}, \citenamefont {Palciauskas}, \citenamefont
  {Rayl}, \citenamefont {Swift}, \citenamefont {Aspnes},\ and\ \citenamefont
  {Kane}}]{article96}%
  \BibitemOpen
  \bibfield  {author} {\bibinfo {author} {\bibfnamefont {Leo~P.}\ \bibnamefont
  {Kadanoff}}, \bibinfo {author} {\bibfnamefont {Wolfgang}\ \bibnamefont
  {G\"otze}}, \bibinfo {author} {\bibfnamefont {David}\ \bibnamefont
  {Hamblen}}, \bibinfo {author} {\bibfnamefont {Robert}\ \bibnamefont {Hecht}},
  \bibinfo {author} {\bibfnamefont {E.~A.~S.}\ \bibnamefont {Lewis}}, \bibinfo
  {author} {\bibfnamefont {V.~V.}\ \bibnamefont {Palciauskas}}, \bibinfo
  {author} {\bibfnamefont {Martin}\ \bibnamefont {Rayl}}, \bibinfo {author}
  {\bibfnamefont {J.}~\bibnamefont {Swift}}, \bibinfo {author} {\bibfnamefont
  {David}\ \bibnamefont {Aspnes}}, \ and\ \bibinfo {author} {\bibfnamefont
  {Joseph}\ \bibnamefont {Kane}},\ }\bibfield  {title} {\enquote {\bibinfo
  {title} {Static phenomena near critical points: Theory and experiment},}\
  }\href {\doibase 10.1103/RevModPhys.39.395} {\bibfield  {journal} {\bibinfo
  {journal} {Rev. Mod. Phys.}\ }\textbf {\bibinfo {volume} {39}},\ \bibinfo
  {pages} {395--431} (\bibinfo {year} {1967})}\BibitemShut {NoStop}%
\end{thebibliography}
\end{document}